# Water evaporation from solute-containing aerosol droplets: effects of internal concentration and diffusivity profiles and onset of crust formation

Majid Rezaei[1,*] and Roland R. Netz[1,†]

[1]*Fachbereich Physik, Freie Universität Berlin, 14195 Berlin, Germany*

**Abstract:** Saliva is primarily composed of water, but additionally includes a variety of organic and inorganic substances such as salt, proteins, peptides, mucins, virions, etc. The presence of such solutes affects the evaporation time of respiratory droplets that are sedimenting in air, and thereby the airborne transmission of infections. From solutions of the coupled heat-conduction and water-diffusion equations within the droplet and in the ambient vapor phase, we find that the solute-induced water vapor-pressure reduction considerably slows down the evaporation process and dominates the solute-concentration dependence of the droplet evaporation time. The evaporation-induced solute-concentration gradient near the droplet surface, which is accounted for using a two-stage evaporation model, is found to further intensify the slowing down of the drying process. On the other hand, the presence of solutes is found to reduce evaporation cooling of the droplet, which causes a slight decrease in the evaporation time. Overall, the first two effects are dominant, meaning that the droplet evaporation time increases in the presence of solutes. The solute-concentration dependence of the water diffusivity inside the droplet does not significantly change the evaporation time. Finally, crust formation on the droplet surface is found to increase the final equilibrium size of the droplet.

[*] m.rezaei@fu-berlin.de
[†] rnetz@physik.fu-berlin.de

**Table 1. List of symbols and notations used**

| | | | |
|---|---|---|---|
| $\tau_{ev}$ | Evaporation time of a pure water droplet | $c_v$ | Water vapor concentration |
| $\tau_{ev}^{sol}$ | Evaporation time of a solute-containing droplet | $c_0$ | Ambient water vapor concentration |
| $\tau^{1st}$ | The time the first drying stage takes | $c_g$ | Saturated water vapor concentration |
| $\tau^{2nd}$ | The time the second drying stage takes | $c_l$ | Liquid water concentration |
| $\Phi$ | Momentary volume fraction of solutes | $c_i$ | Water concentration in the internal core |
| $\Phi_0$ | Initial volume fraction of solutes | $c_s$ | Solute concentration |
| $\Phi_i$ | Volume fraction of solutes in the internal core | $c_s^{max}$ | Solute concentration at the solubility limit |
| $\Phi^{ev}$ | Volume fraction of solutes at the evaporation equilibrium state | $c_s^{ev}$ | Solute concentration at the evaporation equilibrium state |
| $RH$ | Relative humidity | $D_w$ | Water diffusion constant in air |
| $R$ | Droplet radius | $D_w^l$ | Water diffusion constant in pure liquid water |
| $R_0$ | Initial radius of the droplet | $D_w^{sol}$ | Water diffusion constant in water solution |
| $R_{ev}$ | Equilibrium radius of the droplet | $v_w$ | Liquid water molecular volume |
| $R_i$ | Radius of the internal core | $k_e$ | Evaporation reaction rate constant |
| $R°$ | Droplet radius at the end of the 1st drying stage | $k_c$ | Condensation reaction rate constant |
| $R_i°$ | Radius of the internal core at the end of the 1st drying stage | $\Delta T$ | Evaporation induced temperature reduction at the droplet surface |
| $r$ | Radial distance from the droplet center | $T_0$ | Ambient temperature |
| $j$ | Evaporation flux | $T_s$ | Droplet surface temperature |
| $J$ | Total evaporation flux | $\lambda_{air}$ | Heat conductivity of air |
| $J_h$ | Total heat flux into the droplet | $N_A$ | Avogadro constant |
| $h_{ev}$ | Molecular evaporation enthalpy of water | $\gamma$ | Activity coefficient of water |

**Table 2. List of numerical constants used**

| | |
|---|---|
| $D_w$ | $2.5 \times 10^{-5} \ m^2/s$ at 25 °C [1] |
| $D_w^l$ | $2.3 \times 10^{-9} \ m^2/s$ at 25 °C [2] |
| $v_w$ | $3 \times 10^{-29} \ m^3$ at 25 °C [3] |
| $\lambda_{air}$ | $0.026 \ W/mK$ at 25 °C [4] |
| $h_{ev}$ | $7.3 \times 10^{-20} \ J$ at 25 °C [5] |
| $c_g$ | $7.69 \times 10^{23} \ m^{-3}$ at 25 °C and $1.62 \times 10^{23} \ m^{-3}$ at 0 °C [5] |
| $k_c$ | $370 \ m/s$ at 25 °C [6] |

# 1. Introduction

Evaporation from solute-containing droplets is an important process in many industrial and scientific applications ranging from pharmacology, agriculture, food and cosmetics production to medical, biochemical, material, and soil sciences [7-12]. In addition, it plays an important role in the transmission of infectious diseases and respiratory viruses via the airborne route [6, 13-16], which is the main focus of the present study. Water evaporation from a saliva droplet containing non-volatile solutes might eventually produce a droplet nucleus, which is a small light particle at a minimal moisture level that stays floating in air for a long time [17]. The creation of such droplet nuclei can significantly influence the infection risk from virus-containing respiratory droplets, especially in the indoor environment [14, 17], by increasing the sedimentation time of droplets. It is, therefore, of utmost importance to investigate how respiratory droplets dry out and how their evaporation kinetics is affected by different factors such as the droplet composition and its initial size, the ambient relative humidity, temperature, and the surrounding air flow.

Although there are many studies [11, 18-21] concerning the drying process of droplets placed on solid surfaces, also known as sessile droplets, fewer investigations have been conducted on free droplets surrounded by air, which is partly due to a lack of experimental data. Indeed, designing and conducting experiments on free droplets faces many challenges, mostly since such droplets are ephemeral and difficult to follow [8]. Therefore, analytical and numerical modelling methods must be used to better understand the details of the evaporation mechanism that is relevant for sedimenting droplets. Although modelling of free droplets does not face complexities due to substrate-droplet interactions that control the droplet shape at the solid/liquid/air interface [22, 23] and Marangoni flow [24, 25], which occur in the case of sessile droplets, other factors such as evaporation-induced concentration gradients inside the droplet [26] and the possibility of crust formation on the droplet surface [27], which are consequences of the increasing solute surface concentration during evaporation [28], cause difficulties in modelling free droplets in the presence of solutes. In addition, physical and chemical properties of the drying droplets, such as the internal viscosity [29, 30], the diffusivity of solvent and solutes in the liquid phase [30], and the activity coefficient of solvent [31], are dependent on the local concentration of solutes (and, consequently, on both position and time), which makes the problem even more complex. An important challenge in modelling evaporating droplets is, therefore, to account for the concentration gradients created and developed within the droplets during the solvent evaporation process.

There are a few studies that address concentration gradients inside a drying droplet containing solutes [26, 32-34] and propose analytical and numerical methods for modelling the evaporation process before [33-36] and after [27, 35, 37] crust formation. Also, internal concentration gradients have been experimentally captured for drying respiratory fluids suspended on superhydrophobic substrates using optical and fluorescence microscopy [38]. Such concentration gradients are found to affect the evaporation process not only by decreasing the evaporation rate [34], but also by influencing the morphological evolution of the drying droplet [27, 39], which is a determining factor in the shape of the droplet nuclei produced at the end of the evaporation process. An important physical parameter is the ratio of the evaporation rate to the diffusive transport rate of solute particles inside the droplet [40]. When the evaporation process is very slow or, alternatively, the internal diffusion is very fast, the solute particles have enough time to redistribute during the drying process. In such case, the solute particles remain evenly distributed within the droplet to form a full solid particle at the end of the drying process (Fig.1, scenario a). In the case of fast evaporation, however, the solvent evaporation from the droplet surface increases the solute concentration at the surface and creates a concentration gradient in the droplet (see Fig. 1, scenarios b-f). If the evaporation rate is not high enough to increase the solute concentration at the droplet surface up to its solubility limit, the concentration gradient gradually disappears as the evaporation proceeds (due to the decreased evaporation rate) and a full solid particle remains at the end of the drying process (Fig.1, scenario b). When the surface becomes highly populated with solutes, however, two-phase coexistence occurs and the solute particles at the surface form a solid crust (see Fig. 1, scenarios c-e), which is expected to mainly affect the drying mechanism and to considerably slow down the evaporation process [39].

Depending on the type of solutes, the forming crust might be either a dry crust (expected for salty droplets) [8, 41] or a gel-like wet skin consisting of polymers or proteins and other suspended particles [42, 43]. In the presence of a dry crust, the liquid from the internal liquid core reaches the crust surface through capillary action within the crust pores [39, 44]. As the evaporation proceeds, the crust will become thicker and the crust pores will lengthen, which increases the resistance to heat and mass transfer and decreases the evaporation rate. In the case of a wet crust, water evaporation continues via diffusion through the gel crust [42]. Since the diffusion coefficient and the concentration gradient within the gel phase are both much lower than in solution, the evaporation rate is expected to fall after the formation of a wet crust as well. Regardless of the crust type, crust formation is expected to determine the final morphology of the droplet by producing a hollow structure, with a size larger than expected in the absence of crust. Depending on the type of the solute particles within the droplet, the hollow structure produced at the end of the evaporation process may or may not contain small agglomerates of solutes [27] (Fig.1, scenarios c and d). Also, a crust collapse might occur when the crust cannot withstand the pressure difference caused by the continued solvent evaporation[39] (Fig.1, scenario e). Another scenario accounts for non-uniform drying conditions due to amplification of concentration inhomogeneities over the droplet surface, which is probable when the solvent evaporation is very fast. In such case, an irregularly shaped dry particle, also known as wrinkled particle, will form at the end of the drying process [39] (Fig.1, scenario f).

In this paper, a droplet evaporation model is presented that accounts for effects of non-volatile solutes on the drying process. We address evaporation-induced water concentration gradients inside the droplet, the dependence of the internal diffusivity on the solute concentration, the water vapor-pressure reduction in the presence of solutes, and the effect of solutes on the evaporation cooling. We also include the possibility of crust formation in our calculations to evaluate its effect on the droplet size. All our calculations are done in the one-phase regime, meaning from the beginning of the evaporation process up to the point where crust formation sets in at the surface. We find that the presence of non-volatile solutes slows down the droplet evaporation process due to the water vapor-pressure reduction and the presence of solute-induced water concentration gradients inside the droplet. The effect of solutes on the internal water diffusivity is found to play only a minor role in determining the droplet evaporation time, and can therefore be neglected. Crust formation is suggested to affect the morphology of the final droplet nuclei by producing a hollow particle with a size larger than expected in the absence of crust.

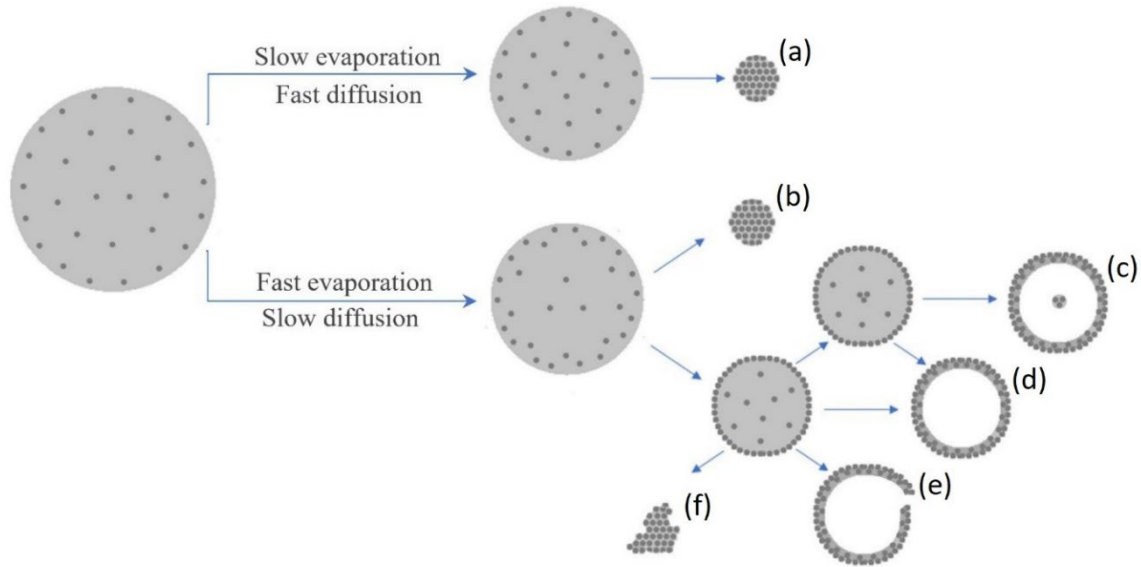

Figure 1 Schematic illustration of possible scenarios for the morphological evolution of a solute-containing droplet during the drying process. When the solvent evaporation is very slow (or the internal particle diffusion inside the droplet is sufficiently rapid), the solute particles have enough time to redistribute in the liquid phase and thus the internal solute concentration remains uniform throughout the evaporation process (except the possibility of two-phase coexistence when the solute-solubility limit is reached). In such case, a solid particle will be produced at the end of the drying process (scenario a). In the case of fast solvent evaporation, however, the shrinkage rate of the droplet radius overcomes the diffusion rate of the solutes in the solution, which leads to an increased solute concentration at the droplet surface. In this case, depending on the type of the solutes and their solubility limit in the solvent, the dry particle produced at the end of the evaportion process might be a solid particle (scenario b) or a hollow particle that may (scenario c) or may not (scenario d) contain small agglomerates of solutes [27]. Alternatively, a collapsed dry particle (scenario e) will be produced when the crust formed at the droplet surface cannot withstand the pressure difference caused by the continued internal solvent evaporation [39]. Very fast solvent evaporation can lead to non-uniform drying conditions due to concentration inhomogeneities of the droplet [39]. In such case, an irregularly shaped dry particle, also known as wrinkled particle, will be formed at the end of the drying process (scenario f).

## 2. Results and discussion

**Droplet evaporation in the presence of non-volatile solutes assuming that the internal water diffusion is sufficiently rapid**

The main objective of the present study is to model water evaporation from a solute-containing droplet that is sedimenting in air. For later comparison with our main results, we first assume that the particle diffusion inside the droplet is sufficiently rapid, so that the water concentration at the droplet surface does not differ significantly from the mean water concentration in the droplet. Also, the effect of solute concentration on evaporation cooling is neglected at first. The radius-dependent evaporation time $t(R)$, which is the time it takes for the droplet radius to decrease from its initial value $R_0$ to $R$, can be approximated from the simultaneous solution of the water concentration and energy diffusion equations as (see reference [6] and Appendix A)

$$t(R) = \frac{R_0^2}{\theta\left(1 - \frac{RH}{\gamma}\right)} \left[1 - \frac{R^2}{R_0^2} - \frac{2R_{ev}^2}{3R_0^2} \ln\left(\frac{R_0(R - R_{ev})}{R(R_0 - R_{ev})}\right)\right] \tag{1}$$

where $R_0$ and $R_{ev}$ are the initial and the evaporation-equilibrium radii of the droplet, respectively, $RH$ denotes the relative humidity, and $\gamma$ is the water activity coefficient accounting for non-ideal effects due to water-solute interactions (more details on $\gamma$ are provided in Appendix B.1). The prefactor $\theta$ in Eq. 1 is given by

$$\theta = 2\gamma D_w c_g v_w \left(\frac{1}{1 + \gamma \varepsilon_C \varepsilon_T}\right) \tag{2}$$

where $D_w$ is the water diffusion constant in air, $c_g$ is the saturated water vapor concentration, and $v_w$ is the water molecular volume in the liquid phase. As discussed in Appendix A, the factor $\frac{1}{1 + \gamma \varepsilon_C \varepsilon_T}$ in Eq. 2 accounts for the evaporation-induced droplet cooling, while the coefficients $\varepsilon_C$ and $\varepsilon_T$ describe the reduction of the water vapor concentration at the droplet surface due to the temperature depression (see Eq. A6) and the dependence of the temperature depression at the droplet surface on the relative humidity and the momentary solute volume fraction (see Eqs. A9 and A10), respectively. For a water droplet at 25°C, the coefficients $\varepsilon_C$ and $\varepsilon_T$ are given by $\varepsilon_c = 0.032\ K^{-1}$ and $\varepsilon_T \equiv \frac{D_w c_g h_{ev}}{\lambda_{air}} = 54$ and, thus, the evaporation-cooling factor is equal to $\sim 0.36$, which indicates that evaporation cooling considerably increases the droplet evaporation time. It should be noted that the evaporation-cooling factor $\theta$ in Eq. 2 neglects the effect of non-volatile solutes (which will be taken into account later), meaning that this factor is derived for a pure water droplet, to make the governing equations analytically solvable (see Appendix A).

An approximate expression for the droplet evaporation time follows by setting $R = R_{ev}$ in Eq. 1, with $R_{ev}$ being the evaporation-equilibrium droplet radius given by (see Eq. A14)

$$R_{ev} = R_0 \left(\frac{\Phi_0}{1 - \frac{RH}{\gamma}}\right)^{1/3} \tag{3}$$

Here, $\Phi_0$ denotes the initial volume fraction of solutes. Neglecting the logarithmic term in equation 1, which reflects the osmotic slowing down of the evaporation due to the water vapor-pressure reduction, the droplet evaporation time $\tau_{ev}^{non-log}$ can be estimated as

$$\tau_{ev}^{non-log} = \tau_{ev}\left(1 - \frac{R_{ev}^2}{R_0^2}\right) \tag{4}$$

where $\tau_{ev}$ is the evaporation time in the absence of solutes [6]

$$\tau_{ev} = \frac{R_0^2}{\theta\left(1 - \frac{RH}{\gamma}\right)} \tag{5}$$

As mentioned above, equation 4 is derived neglecting the water vapor-pressure reduction during the evaporation process as well as the concentration dependence of evaporation cooling. The importance of these factors is evaluated next.

**Effect of the water vapor-pressure reduction**

To account for the solute-induced water vapor-pressure reduction, the logarithmic term in equation 1 is now included. Since near the evaporation equilibrium, the droplet radius varies very slowly with time, as follows from the logarithmic term in Eq. 1 (see Fig. 2a), we can no longer use the mentioned definition for evaporation time (i.e., the time at which $R = R_{ev}$) that leads to Eq. 4. Instead, we define the evaporation time as the time at which the equilibrium droplet radius $R_{ev}$ is 99% of the droplet radius $R$

$$\tau_{ev} = t\left(\frac{R_{ev}}{0.99}\right) \tag{6}$$

From Eq. 1 and considering the definition provided by Eq. 6, the evaporation time in the presence of solutes follows as

$$\tau_{ev}^{sol} = \tau_{ev}^{non-log} + \tau_{ev}^{log} = \tau_{ev}\left(1 - \frac{R_{ev}^2}{R_0^2}\right) - \frac{2R_{ev}^2}{3R_0^2}\tau_{ev}\ln\left(\frac{0.01}{1 - \frac{R_{ev}}{R_0}}\right) \tag{7}$$

$$= \tau_{ev}\left(1 + \frac{2R_{ev}^2}{3R_0^2}\left(3.105 + \ln\left(1 - \frac{R_{ev}}{R_0}\right)\right)\right)$$

The logarithmic term $\tau_{ev}^{log}$ in Eq. 7 reflects the increase in the evaporation time due to the solute-induced water vapor-pressure reduction

$$\tau_{ev}^{log} = \frac{2R_{ev}^2}{3R_0^2}\tau_{ev}\left(4.605 + \ln\left(1 - \frac{R_{ev}}{R_0}\right)\right) \tag{8}$$

Figure 2b shows the evaporation time obtained from equation 7 along with its logarithmic and non-logarithmic parts as a function of the initial volume fraction of solutes $\Phi_0$. This figure indicates that at very low values of $\Phi_0$, $\tau_{ev}^{log}$ is negligible compared to $\tau_{ev}^{non-log}$, meaning that equation 4 is quite accurate. An increase in $\Phi_0$, however, intensifies the effect of the solute-induced water vapor-pressure reduction and thus increases $\tau_{ev}^{log}$, while nevertheless $\tau_{ev}^{non-log}$ decreases with $\Phi_0$ due to the increased equilibrium radius of the droplet (see equations 3 and 4). In the case of droplets with high initial volume fractions of solutes $\Phi_0$, therefore, the logarithmic terms in equations 1 and 7 are important and can no longer be neglected. By equating equations 4 and 8, one finds that the effect of the water vapor-pressure reduction becomes dominant when $\Phi_0$ becomes higher than $\sim 0.13(1 - RH/\gamma)$.

Figure 2b also shows that an increase in $\Phi_0$ causes a non-monotonic variation in the evaporation time, which cannot be observed when the water vapor-pressure reduction is neglected. By differentiating equation 7 with respect to $\Phi_0$, the initial volume fraction of solutes at which the droplet experiences the maximum evaporation time can be estimated as

$$\Phi_0^{\tau_{ev,max}} \simeq 0.55 \left(1 - \frac{RH}{\gamma}\right) = 0.55 \Phi_{ev} \tag{9}$$

with $\Phi_{ev} = 1 - \frac{RH}{\gamma}$ being the volume fraction of solutes at evaporation equilibrium, i.e., when the evaporation flux (given by Eq. A11) vanishes.

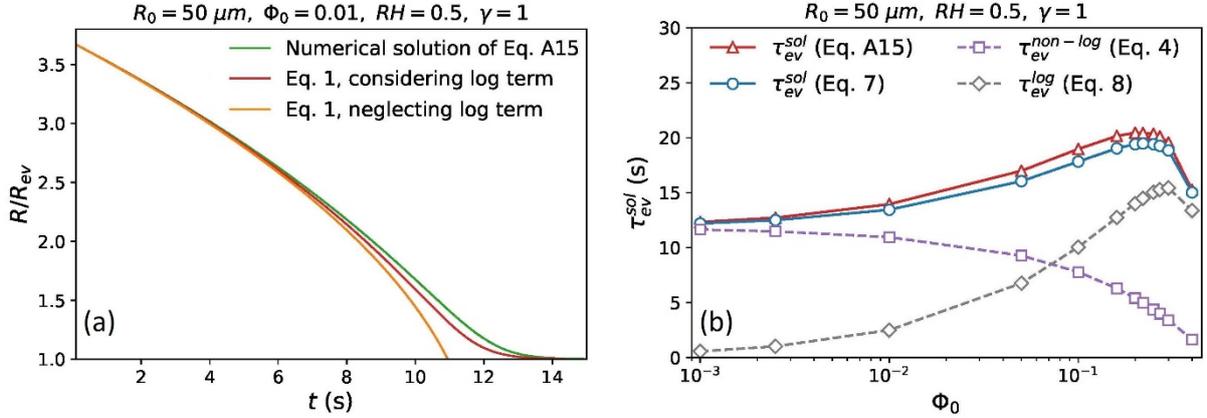

Figure 2 Effect of the solute-induced water vapor-pressure reduction on the drying process in the limit of an infinitely large internal water diffusion constant (the solute-concentration dependence of evaporation cooling is neglected here). (a) Variation of the droplet radius $R$ with time $t$ for a droplet with initial radius of $R_0 = 50 \, \mu m$ for initial solute volume fraction $\Phi_0 = 0.01$ and $RH = 0.5$. The green line shows the results obtained from numerical solution of the differential equation A15, which is derived from the mass conservation of the droplet (see Appendix A). Red and orange lines indicate the results estimated from Eq. 1, which is the approximate analytical solution of Eq. A15, with and without considering the logarithmic term that reflects the water vapor-pressure reduction. (b) Evaporation time $\tau_{ev}^{sol}$ as a function of the initial volume fraction of solutes $\Phi_0$. Solid lines show the evaporation times calculated from equations A15 and 7 at $RH = 0.5$ and broken lines indicate the non-logarithmic and the logarithmic contributions to the evaporation time, which are obtained from equations 4 and 8, respectively.

**Effect of non-volatile solutes on evaporation cooling**

The evaporation-cooling factor $1/(1 + \gamma \varepsilon_C \varepsilon_T)$ in Eq. 2 is obtained for a pure water droplet. The presence of non-volatile solutes affects this term by decreasing the saturated water vapor concentration $c_g$ and, consequently, the prefactor $\varepsilon_T \equiv \frac{D_w c_g h_{ev}}{\lambda_{air}}$ (see Appendix A for more details). Accounting for these solute effects, the evaporation-induced temperature reduction at the surface of a solute-containing droplet is obtained as

$$\Delta T = \gamma \varepsilon_T \left( \frac{1 - \frac{\Phi_0 R_0^3}{R^3} - \frac{RH}{\gamma}}{1 + \gamma \varepsilon_c \varepsilon_T \left(1 - \frac{\Phi_0 R_0^3}{R^3}\right)} \right) \qquad (10)$$

Equation 10 reveals that not only $\Delta T$ is linearly related to the relative humidity, as observed in the absence of solutes [6], but it also depends on the momentary volume fraction of solutes $\Phi_0(R_0^3/R^3)$, which itself is time dependent. This equation clearly shows that at evaporation equilibrium, i.e., when the momentary volume fraction of solutes reaches $1 - RH/\gamma$ and thus the evaporation flux vanishes (see Eq. A11), $\Delta T$ goes to zero, meaning that the droplet reaches thermal equilibrium with its environment as well (see Fig. 3a).

Considering the concentration-dependent temperature depression at the droplet surface, which is given by Eq. 10, one can rewrite the prefactor $\theta$ in Eq. 2 in the presence of solutes as (see Eq. A21)

$$\theta^{sol} = \frac{2\gamma D_w c_g v_w}{1 + \gamma \varepsilon_c \varepsilon_T \left(1 - \frac{\Phi_0 R_0^3}{R^3}\right)} \qquad (11)$$

and repeat the calculations to rederive the evaporation time. As detailed in Appendix A, the evaporation time is determined by the condition of total mass conservation of the droplet, which gives rise to the differential equation A13 that is not analytically solvable. Instead, this equation is numerically solved to yield $R(t)$. Figure 3b shows the time-dependent droplet radii calculated from equations A13 and A15, which are derived with and without considering the concentration dependence of the evaporation-cooling effect, respectively. This figure indicates that the solute-induced reduction of $\Delta T$ slightly speeds up the evaporation process. As expected, this speed up becomes more significant as the evaporation proceeds because of the increased solute concentration within the droplet. Also, the solute effect on evaporation cooling is found more pronounced for droplets with higher initial solute volume fractions (see Fig. 3b and its inset).

Using the definition provided by Eq. 6, the evaporation time is calculated from numerical solution of equations A13 and A15 (i.e., with and without considering the effect of solutes on $\Delta T$). The results are shown by solid and broken lines in Fig. 3c. This figure also shows the results obtained from equation 7 (dotted lines), which is an approximate analytical solution of equation A15 (see Appendix A). This figure clearly shows that the solute-concentration dependence of $\Delta T$ causes a decrease in the evaporation time, as discussed above. The approximation used in derivation of equation 7 (see Eq. A19) is, however, found to partly compensate the difference caused by neglecting such concentration dependence, especially at high relative humidities. According to this figure, when the initial volume fraction of solutes is equal to or less than $\Phi = 0.01$, which is reported as a normal volume fraction of solutes in saliva droplets [45], one can safely use equation 7 to calculate $\tau_{ev}$. At higher values of $\Phi_0$, however, this approximation starts to become inaccurate. Figure 3c also shows that the initial volume fraction of solutes at which the droplet exhibits a maximum in the evaporation time decreases with $RH$, as can be seen from equation 9.

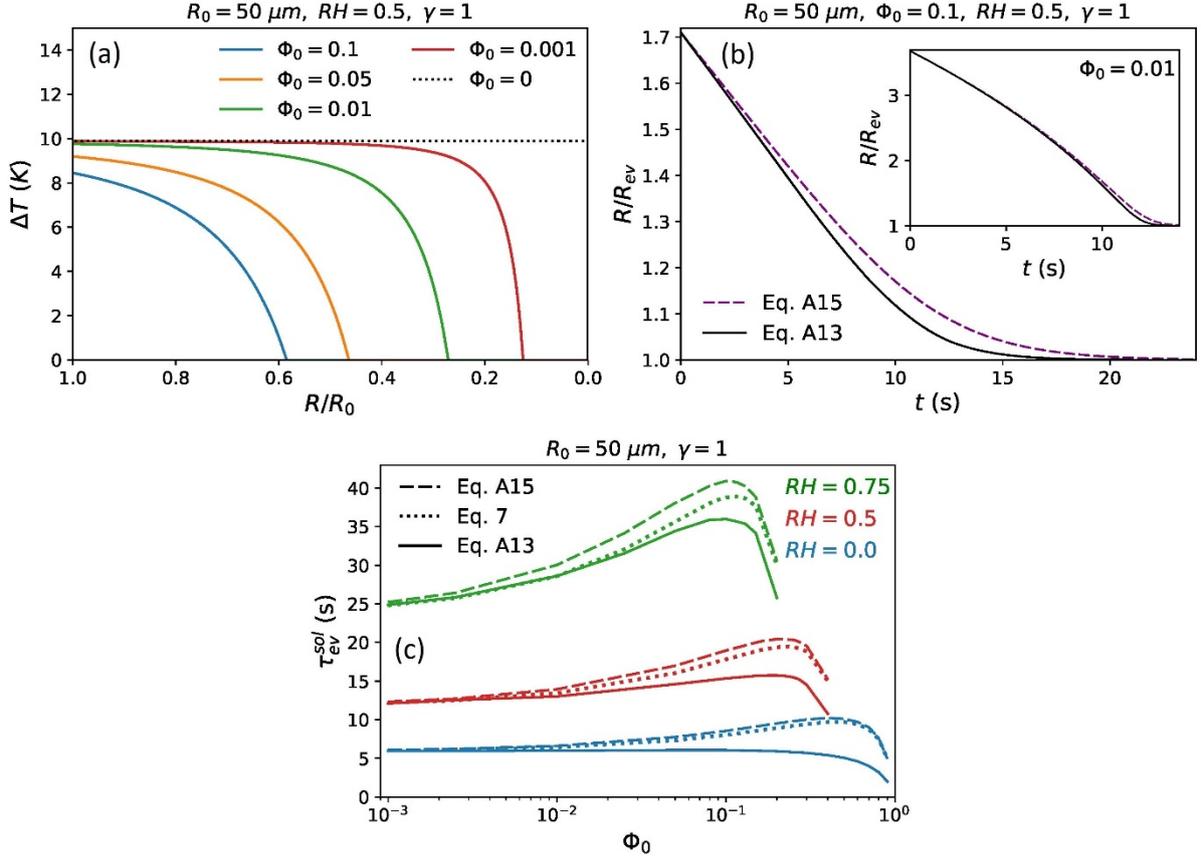

Figure 3 Effect of non-volatile solutes on evaporation cooling and, consequently, on the evaporation time (the results are shown in the limit of infinitely high internal water diffusivity). (a) Evaporation-induced temperature reduction at the droplet surface, $\Delta T$, as a function of the droplet radius $R$. The results are obtained from Eq. 10 for fixed relative humidity of $RH = 0.5$ and different initial solute volume fractions $\Phi_0$. The dotted line indicates the results neglecting the solute-concentration dependence of evaporation cooling, i.e., for $\Phi_0 = 0$. (b) Variation of the droplet radius $R$ with time $t$. The results are shown for $RH = 0.5$ and for two different initial solute volume fractions $\Phi_0 = 0.1$ (shown in the main figure) and $\Phi_0 = 0.01$ (shown in the inset). Solid lines indicate the results from numerical solution of Eq. A13, which account for the solute-concentration dependence of evaporation cooling, and broken lines show the results from numerical solution of Eq. A15 that is derived neglecting such solute-concentration dependence. (c) Evaporation time $\tau_{ev}^{sol}$ as a function of initial solute volume fraction $\Phi_0$. Solid and broken lines indicate the results obtained from numerical solution of equations A13 and A15, respectively, and dotted lines show the Eq. 7 (which results from the approximate analytical solution of Eq. A15).

**Droplet evaporation in the presence of non-volatile solutes considering evaporation-induced internal concentration gradients**

In the previous section, the particle diffusivity inside the droplet was assumed to be sufficiently rapid, so that the solute particles remain evenly distributed throughout the liquid phase. In the fast evaporation scenario (see Fig. 1), however, the droplet experiences an increased solute concentration at the droplet surface. To account for the resulting water concentration gradient in the liquid phase, one needs to solve the diffusion equation not only outside, but also inside the droplet

$$\frac{dc_l(r,t)}{dt} = \frac{1}{r^2}\frac{\partial}{\partial r}\left(D_w^l[c_l(r,t)]r^2\frac{\partial c_l(r,t)}{\partial r}\right) \qquad (12)$$

Here, $c_l(r,t)$ is the liquid water concentration profile inside the droplet and $D_w^l[c_l(r,t)]$ is the water diffusion coefficient in liquid water, which in general depends on the water concentration and, consequently, on time and position. Equation 12 has a singularity at the center of the droplet, which is preempted because in the model used here, the droplet volume is divided into two regions: an internal core, where the concentration profile remains uniform, and an outer shell, where the liquid phase experiences a concentration gradient (see Fig. 4). To calculate the water concentration profile inside the droplet, therefore, it is sufficient to solve the diffusion equation (Eq. 12) in the outer shell. Here, we consider a two-stage evaporation model. In the first stage (see Fig. 4a), the outer shell grows towards the droplet core. In this stage, the radius of the internal core $R_i$ shrinks with time while the water concentration in the internal core $c_i$ remains equal to its initial value. When $R_i$ reaches a cut off value $R_i^\circ$ (which is of molecular size), the drying process turns into the second stage (Fig. 4b). In this stage, the internal core radius $R_i$ is fixed to $R_i^\circ$ while $c_i$ is considered as a time-dependent parameter (which is expected to reach the equilibrium concentration at the end of the second drying stage).

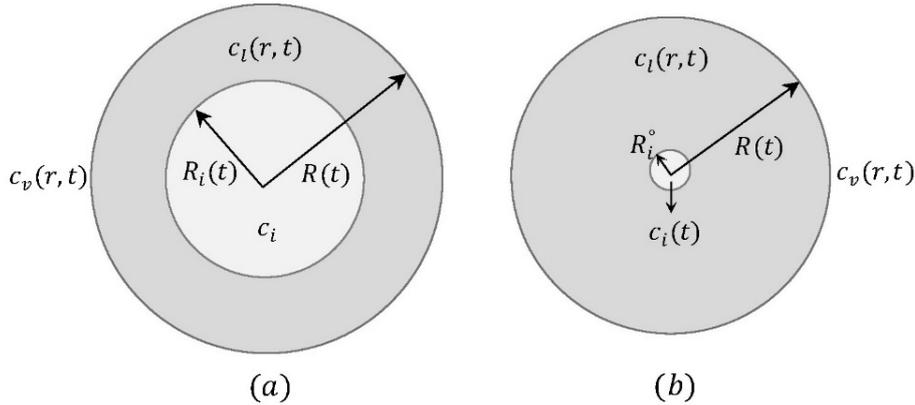

Figure 4 The model used to account for the evaporation-induced water concentration gradient at the droplet surface. Here, the water concentration profile within the internal core of radius $R_i(t)$ (the bright area) is assumed to remain uniform throughout the drying process $c_l = c_i$ while the outer shell of radius $R(t)$ (the dark area) exhibits a water concentration gradient $c_l = c_l(r)$. This model leads to two drying stages: (a) in the first stage, the internal concentration $c_i$ is kept constant while the radius of the internal core $R_i$ varies with time, (b) when the radius of the internal core reaches a very small value of $R_i^\circ$ (which is of molecular size) the drying process turns into the second stage, where $c_i$ is time-dependent and $R_i$ is fixed to $R_i^\circ$.

We first assume that the internal water diffusion coefficient $D_w^l$ is independent of solute concentration (concentration dependence of $D_w^l$ is discussed next). Using the two-stage model described above, the liquid water density profiles in the first drying stage $c_l^{1st}(r)$ and in the second drying stage $c_l^{2nd}(r)$ are given as (the calculations are detailed in Appendix B)

$$c_l^{1st}(r) = \begin{cases} \dfrac{1-\Phi_0}{v_w} & r \leq R_i \\[1em] \dfrac{1-\Phi_0}{v_w} + \dfrac{\left(1-\Phi_0 - \dfrac{RH}{\gamma}\right)\left(\dfrac{R}{r} - \dfrac{R}{R_i}\right)}{v_w\left(\dfrac{\alpha}{\gamma} - 1 + \dfrac{R}{R_i} + \alpha\varepsilon_T\varepsilon_c(1-\Phi_0)\right)} & R_i < r \leq R \end{cases} \tag{13}$$

$$c_l^{2nd}(r) = \begin{cases} \dfrac{1 - \dfrac{\Phi_0 R_0^3}{R^3} + \mu\dfrac{RH}{\gamma}}{v_w(1+\mu)} & r \leq R_i^\circ \\[1em] \dfrac{1 - \dfrac{\Phi_0 R_0^3}{R^3} + \mu\dfrac{RH}{\gamma}}{v_w(1+\mu)} + \dfrac{\dfrac{1}{v_w}\left(1 - \dfrac{\Phi_0 R_0^3}{R^3} - \dfrac{RH}{\gamma}\right)\left(\dfrac{R}{r} - \dfrac{R}{R_i^\circ}\right)}{\left(\dfrac{\alpha}{\gamma} - 1 + \dfrac{R}{R_i^\circ}\right)(1+\mu) + \alpha\varepsilon_T\varepsilon_c\left(1 - \dfrac{\Phi_0 R_0^3}{R^3} + \mu\dfrac{RH}{\gamma}\right)} & R_i^\circ < r \leq R \end{cases} \tag{14}$$

$R$ and $R_i$ in the above equations are time dependent and $\mu$ in equation 14 is a function of $R$, which is given by

$$\mu = -\dfrac{\dfrac{R_i^{\circ 2}}{R^2} + 2\dfrac{R}{R_i^\circ} - 3}{2\left(\dfrac{\alpha}{\gamma} - 1 + \dfrac{R}{R_i^\circ} + \alpha\varepsilon_T\varepsilon_c\right)} \tag{15}$$

The numerical prefactor $\alpha$ in equations 13-15 is given by $\alpha = D_w^l/(D_w c_g v_w)$ and describes the ratio of internal and external water diffusivities. Considering the values listed in table 2, this prefactor is approximately equal to $\alpha \approx 4$ at 25°C. Using the assumption of infinitely rapid diffusion in the droplet, $\alpha$ goes to infinity. In the present study, the calculations are done for different values of $\alpha$ to investigate the effect of the internal water diffusion constant on the water evaporation process.

Neglecting the probability of crust formation due to the increased solute concentration at the droplet surface (which is discussed later), one can calculate the evaporation time as

$$\tau_{ev} = \tau^{1st} + \tau^{2nd} \tag{16}$$

where $\tau^{1st}$ and $\tau^{2nd}$ are the times of the first and the second drying stages, respectively. In the present study, $\tau^{1st}$ and $\tau^{2nd}$ are obtained from equations B32 and B39, using the numerical approaches elaborated in Appendix B. In our calculations, $R_i^\circ$ (i.e., the radius of the internal core in the second drying stage) is set to $1\ nm$, which is small enough so that the calculated evaporation time is independent of $R_i^\circ$ (as demonstrated in Fig. 5a). Figure 5b shows the resulting evaporation time according to Eq. 16 as a function of the internal water diffusion constant for a droplet with an initial radius of $R_0 = 50\ \mu m$. This figure demonstrates that an increase in $D_w^l$ decreases the evaporation time, as expected. At high internal diffusion coefficients, however, $\tau_{ev}$ becomes independent of $D_w^l$. Figure 5b also confirms that for $\Phi_0 =$ 0.01, equation 7 accurately predicts the evaporation time in the limit of $D_w^l \to \infty$, as discussed in the previous section.

The dotted vertical line in Fig. 5b indicates the value reported [2] for the water diffusion coefficient in pure water at 25°C $D_w^{l,25°C} = 2.3 \times 10^{-9} \ m^2/s$ (see table 2). Neglecting the concentration dependence of the diffusion coefficient, this value can be used to estimate the evaporation time for a solute-containing water droplet at 25°C. Figure 5c shows the resulting evaporation times together with those calculated for $D_w^l \to \infty$, i.e., the results obtained assuming that the internal concentration gradients are infinitely small. According to this figure, although the effect of internal concentration gradients is negligible at low initial volume fractions of solutes, neglecting this effect can make a relatively large error at higher values of $\Phi_0$, especially when $\Phi_0$ becomes close to $\Phi_0^{\tau_{ev,max}}$ (see Eq. 9). The relative error is found to be more significant in dry environments because the drier the air, the faster the water evaporates from the droplet surface and the larger the concentration gradient is.

As mentioned before, the differential equations B32 and B39 that are used here to calculate the evaporation time are not analytically solvable and, hence, the evaporation time is calculated using numerical methods. To provide an equation for estimating $\tau_{ev}$ as a function of the relevant parameters considering all the factors that have been discussed so far (i.e., solute-induced water vapor-pressure reduction, concentration dependence of the evaporation-cooling effect, and the evaporation-induced concentration gradients inside the droplet), we construct a heuristic fit function inspired by equation 7

$$\tau_{ev} = a_{\tau 1} \frac{R_0^2}{\theta' \left(1 - \frac{RH}{\gamma}\right)} \left[1 + a_{\tau 2} \left(\frac{R_{ev}}{R_0}\right)^2 \left(3.105 + \ln\left(1 - \frac{R_{ev}}{R_0}\right)\right)\right] \tag{17}$$

where $a_{\tau 1}$ and $a_{\tau 2}$ are fit parameters and $\theta'$ is a modified numerical prefactor given by

$$\theta' = \frac{2\gamma D_w c_g v_w}{1 + \gamma \varepsilon_C \varepsilon_T (1 - \Phi_0)} \tag{18}$$

Figure 5c demonstrates that Eq. 17 (shown by broken lines) with fit parameters $a_{\tau 1} = 1.03$ and $a_{\tau 2} = \frac{5}{6}$ perfectly fits the data over a wide range of $\Phi_0$ and $RH$. It is worth mentioning that the values used for $a_{\tau 1}$ and $a_{\tau 2}$ are the averages of the fit parameters obtained at different conditions (i.e., at different values of $\Phi_0$, $RH$, and $\gamma$).

Finally, the time-dependent variation of the droplet radius during the drying process is assessed. The point to note here is that in the case of a droplet with a finite internal water diffusivity, the first drying stage is dominant. Indeed, when $R_i^\circ$ is set to a sufficiently small value (in the order of a few nano-meters or smaller), the time of the second drying stage $\tau^{2nd}$ goes to zero (as demonstrated in Fig. 5a), meaning that the droplet does not experience the second drying stage. Therefore, the time-dependent radius of a droplet with finite $D_w^l$ can be calculated from equation B32, which describes the first drying stage. Also, $R(t)$ in the limit of $D_w^l \to \infty$ can be obtained from Eq. A13, which neglects the internal water concentration gradients. The results are shown in figure 5d. This figure indicates that in the beginning of the evaporation process, when the evaporation-induced concentration gradients are not significant yet, the shrinkage rate of the droplet radius is almost independent of the internal water diffusion constant. As the evaporation proceeds, however, this parameter becomes important and affects the droplet evaporation time, as expected.

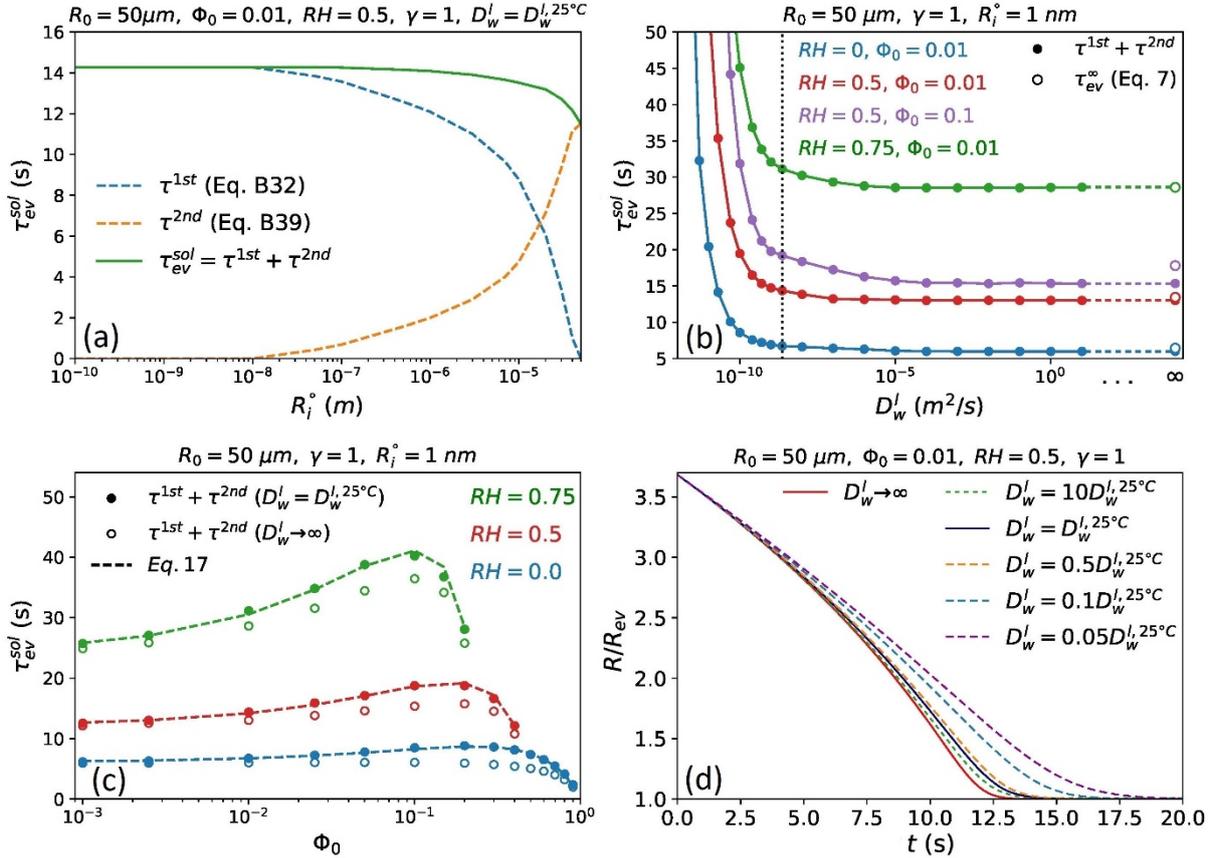

Figure 5 Results obtained using the two-stage evaporation model that accounts for the internal water concentration gradient. (a) The times of the two drying stages ($\tau^{1st}$ and $\tau^{2nd}$) and the total evaporation time $\tau_{ev}^{sol} = \tau^{1st} + \tau^{2nd}$ as a function of $R_i^\circ$, with $R_i^\circ$ being the radius of the internal core $R_i$ in the second drying stage (see Fig. 3). Here, $\tau^{1st}$ and $\tau^{2nd}$ are calculated from equations B32 and B39, respectively, for $RH = 0.5$, $\Phi_0 = 0.01$, and considering the value reported [2] for the water diffusion coefficient at $25°C$ $D_w^{l,25°C} = 2.3 \times 10^{-9} \, m^2/s$ (see table 2). (b) Evaporation time $\tau_{ev}^{sol}$ as a function of the water diffusion coefficient in the liquid phase $D_w^l$. Solid symbols show the results from equations B32 and B39 using $R_i^\circ = 1 \, nm$. The evaporation times estimated from Eq. 7 in the limit of infinitely large internal water diffusion constant are depicted by open symbols to the right. The vertical dotted line indicates $D_w^l = D_w^{l,25°C}$ (see table 2). (c) Variation of the evaporation time $\tau_{ev}^{sol}$ with the initial solute volume fraction $\Phi_0$. Solid and open symbols depict the results for $D_w^l = D_w^{l,25°C}$ and $D_w^l \to \infty$, respectively, which are calculated from equations B32 and B39 (again, using $R_i^\circ = 1 \, nm$). Broken lines indicate the results estimated from Eq. 17 using the fit parameters $a_{\tau 1} = 1.03$ and $a_{\tau 2} = 5/6$. (d) Droplet radius $R$ as a function of time $t$. The results for $D_w^l \to \infty$ are calculated from Eq. A13 and those for $D_w^l < \infty$ are obtained from Eq. B32 for $RH = 0.5$ and $\Phi_0 = 0.01$.

**Solute-concentration dependence of the water diffusion constant in the droplet without crust formation**

The diffusion coefficient of water is known [30] to decrease in the presence of strongly hydrated solutes, such as NaCl, NaI, and NaBr. In contrast, the presence of weakly hydrated solutes, such as KCl, KBr, and KI, increases the water self-diffusion coefficient, presumably by disrupting hydrogen bonding. According to figure 5b, such variations in the water diffusion coefficient slow down or speed up the droplet evaporation process. Therefore, the evaporation model introduced in the previous section must be modified to account for the solute-concentration dependence of water diffusivity inside the droplet. Here, we assume that the internal water diffusion coefficient is linearly dependent on the solute concentration, which approximates the behavior of NaCl and NaBr salts rather accurately [30]. Accordingly, we write

$$D_w^{sol}(r,t) = D_w^l(1 - \beta c_s(r,t)) = D_w^l\left(1 - \frac{\beta}{v_s}(1 - v_w c_l(r,t))\right) \qquad (19)$$

with $D_w^{sol}$ and $D_w^l$ being the water diffusion coefficients in an aqueous solution and in pure water, respectively, $\beta$ being an ion-specific constant ($\sim 0.065\ M^{-1}$ for NaCl and $\sim 0.058\ M^{-1}$ for NaBr solution [30]), $v_s$ being the molecular volume of solutes, and $c_s(r,t) = \frac{1}{v_s}(1 - v_w c_l(r,t))$ being the solute concentration profile. By incorporating equation 19 into the diffusion equation (Eq. 12) and after using the numerical methods described in Appendix C to solve this equation along with the energy diffusion and the mass conservation equations, the time-dependent radius of the evaporating droplet is determined.

According to equation 19, when the solute concentration at the droplet surface reaches a critical value of $c_s^* = \beta^{-1}$, the water diffusion coefficient at the surface of the droplet $D_w^{sol}(R)$ reaches zero. Therefore, in the cases where $c_s^*$ is lower than the equilibrium concentration of solutes $c_s^{ev} = \frac{1-RH/\gamma}{N_A v_s}$, i.e., the solute concentration at which the evaporation rate equals the condensation rate (see Eq. B17), the water diffusion flux at the surface $-D_w^{sol}(R)\frac{\partial c_l}{\partial r}\Big|_{r=R}$ goes to zero before the evaporation flux $k_e c_l(R) - k_c c_v(R)$ vanishes. In such case, the reactive boundary condition given by Eq. B7 cannot be satisfied, which prevents the numerical solution of the diffusion equation from converging. To avoid this problem, we exclusively consider the parameter range of $\beta$ and $RH$ so that $c_s^{ev} \leq c_s^*$. In this limit, the water diffusion coefficient remains non-zero during the entire drying process and minimally reaches zero in the evaporation-equilibrium state. At first, the possibility of phase separation at high solute concentrations is neglected. In reality, when the solute concentration at the droplet surface exceeds the solubility limit, which is normally reached before $D_w^{sol}$ goes to zero, a solid crust forms at the droplet surface. For example, the solubility limit of NaCl in water is around 6.1 M [46], and $\beta$ for a NaCl solution is $\sim 0.065\ M^{-1}$ [30]. Therefore, according to Eq. 19, the water diffusion constant in a liquid droplet of NaCl solution at the onset of crust formation is $\sim 60\%$ of the water diffusion constant in pure water. The possibility of crust formation is accounted for in the next section.

Panels a-c of figure 6 show the variation of the droplet radius with time in the presence of solutes that increase or decrease the water diffusivity. This figure indicates that the concentration dependence of the diffusion coefficient does not significantly change the time-dependent shrinkage of the droplet radius, even for droplet radii close to the final equilibrium radius, where the solute concentration at the droplet surface tends to its maximum value. To find out the reason for this, the variation of the water evaporation flux in the presence of solutes is examined. Our results reveal that not only an increase in $\beta$ decreases the internal water diffusion coefficient at the droplet surface $D_w^{sol}(R)$ (see Fig. 6, panels d-f), as follows from Eq. 19, but it also reduces the growth rate of the outer shell, i.e., the shell where the liquid phase exhibits a concentration gradient (see Fig. 6, panels g-i). The surface concentration of solutes, however, does not change much in the presence of solutes with different values of $\beta$ (see Fig. 6, insets of panels d-f), meaning that the water concentration gradient at the droplet surface $c'(R) = -\frac{\partial c_l}{\partial r}\big|_{r=R}$ increases with increasing $\beta$ (see Fig. 6, insets of panels g-i). As a result, the water evaporation flux $J = 4\pi R^2 D_w^{sol}(R) c'(R)$ is found to remain almost independent of $\beta$ (see Fig. 6, panels j-l). This implies that the effect of the concentration dependence of the water diffusivity inside the droplet on the water evaporation rate and, consequently, on the time-dependent droplet size, is negligible. Therefore, one can safely use equation 17 to estimate the droplet evaporation time in the presence of solutes (in the absence of the possibility of crust formation).

**Effect of the solubility limit of solutes and crust formation**

The results presented so far are obtained without accounting for the possibility of crust formation. Depending on evaporation rate, temperature, and type of the solutes, a solid crust can form, which dramatically affects the drying process and the time it takes for the droplet to reach the evaporation equilibrium state. It is, therefore, of importance to determine when a crust forms on the droplet surface and how it affects the evaporation. It is also of interest to examine whether the concentration-dependent variation of the internal water diffusivity delays or accelerates the crust formation.

Here, we assume that a solid crust forms instantaneously when the solute concentration at the droplet surface $c_s(R)$ reaches the solubility limit $c_s^{max}$. In the present study, the solubility limit of NaCl in water (6.1 M [46]) is used for $c_s^{max}$. The crust can form only if $c_s^{max}$ is lower than the equilibrium solute concentration $c_s^{ev} = \frac{1-RH/\gamma}{N_A v_s}$. The threshold relative humidity below which crust formation can occur during the drying process is, therefore, given by

$$RH^* = \gamma(1 - c_s^{max} v_s N_A) \qquad (20)$$

where $N_A$ is the Avogadro constant and $c_s^{max} v_s N_A$ expresses the maximum possible volume fraction of solutes in the liquid phase (~ 0.11 for a NaCl solution). In the case of droplets containing a NaCl solution, therefore, crust formation is a very probable phenomenon that occurs when the relative humidity is lower than $RH^* \approx 0.89$.

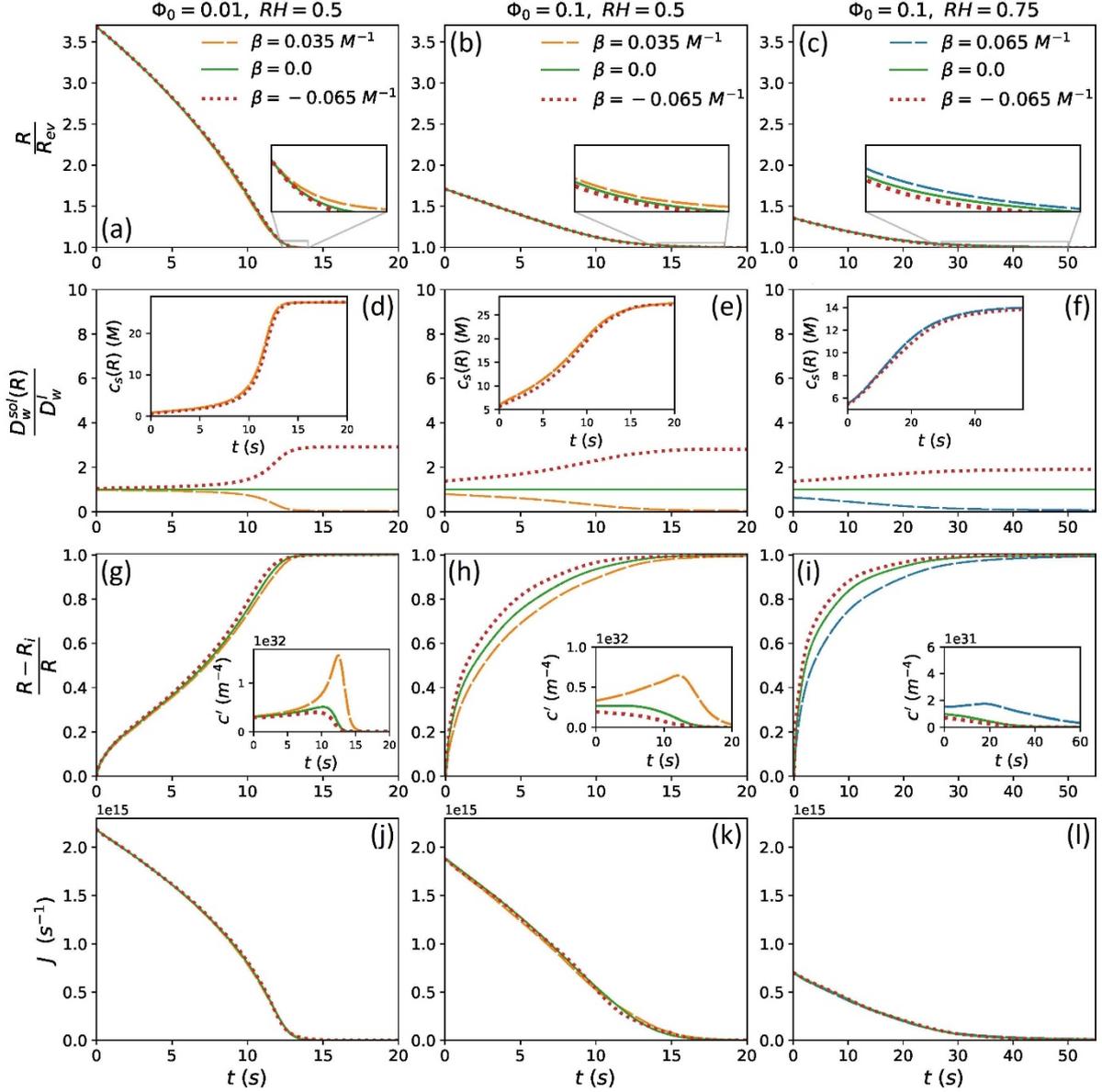

Figure 6 Time-dependent variation of (a-c) the droplet radius $R$, (d-f) the water diffusion constant at the droplet surface $D_w^{sol}(R)$, (d-f, insets) the surface concentration of solutes $c_s(R)$, (g-i) the thickness of the outer droplet shell $R - R_i$, (g-i, insets) the water concentration gradient at the droplet surface $c'(R) = -\partial c/\partial r|_{r=R}$, and (j-l) the total water evaporation flux $J$. The results are obtained using the numerical method described in Appendix C.2, which accounts for the solute-concentration dependence of the internal water-diffusion coefficient, for $\Phi_0 = 0.01$ and $RH = 0.5$ (first column), $\Phi_0 = 0.1$ and $RH = 0.5$ (second column), and $\Phi_0 = 0.1$ and $RH = 0.75$ (third column). Here, it is assumed that the solute and water molecular volumes are equal $v_s = v_w$ and the values reported in table 2 are used for the molecular volume of water $v_w = 3 \times 10^{-29} \, m^3$ and the water diffusion coefficient in pure water $D_w^l = D_w^{l,25°C} = 2.3 \times 10^{-9} \, m^2/s$ (see Eq. 19).

Another assumption made here is that the outer radius of the droplet remains constant in the presence of crust, meaning that the possibility of crust collapse (Fig. 1, scenario e) or wrinkling (Fig. 1, scenario f) is ignored. Accordingly, one can define the droplet equilibrium radius $R_{ev}$ as the smaller of the radii at which the droplet reaches the evaporation equilibrium state (in the absence of crust) and that at which a crust first forms on the droplet surface. Neglecting the concentration dependence of the water diffusivity within the droplet, $R_{ev}$ follows from equation B30 as

$$R_{ev} = R_0 \left( 1 + \frac{\left(\left(\frac{R_i^*}{R_{ev}}\right)^3 - 3\left(\frac{R_i^*}{R_{ev}}\right) + 2\right)\left(1 - \Phi_0 - \frac{RH}{\gamma}\right)}{2\Phi_0 \left(\left(\frac{\alpha}{\gamma} - 1 + \alpha \varepsilon_T \varepsilon_c (1 - \Phi_0)\right)\left(\frac{R_i^*}{R_{ev}}\right) + 1\right)} \right)^{-\frac{1}{3}} \qquad (21)$$

where $R_i^*$ is the radius of the internal core at the moment when the droplet reaches its final equilibrium size $R_{ev}$. According to the water concentration profile Eq. 13, the ratio $R_i^*/R_{ev}$ is given by

$$\frac{R_i^*}{R_{ev}} = \left( 1 - \frac{\frac{\alpha}{\gamma}(\Phi_0 - \Phi_{max})(1 + \gamma \varepsilon_T \varepsilon_c (1 - \Phi_0))}{1 - \Phi_{max} - \frac{RH}{\gamma}} \right)^{-1} \qquad (22)$$

with $\Phi_{max}$ being the maximum possible volume fraction of solutes, i.e., the smaller of $c_s^{max} v_s N_A$ and $1 - RH/\gamma$. In the case when the drying process ends before a crust forms (i.e., when $1 - RH/\gamma < c_s^{max} v_s N_A$), the above equations yield $R_i^* = 0$ and, consequently, $R_{ev} = R_0 \left(\frac{\Phi_0}{1-RH/\gamma}\right)^{1/3}$, which is equal to what we previously obtained in the absence of crust formation (see Eq. 3).

According to figure 7a, when $RH/\gamma$ is lower than the threshold value given by Eq. 20, a crust forms at the surface of the droplet and thus the droplet equilibrium size calculated from Eq. 21 is greater than what would be expected in the absence of a crust. In this case, an increase in $RH/\gamma$ slightly decreases $R_{ev}$, which can be attributed to the reduced water evaporation rate in humid environment or in water solutions with low water-activity coefficients. Indeed, when the evaporation rate decreases, the solute particles at the droplet surface have more time for diffusional motion towards the center of the droplet and thereby, the surface concentration rises more slowly. This in turn causes a slight delay in the crust formation and leads to a smaller droplet equilibrium size. When $RH/\gamma$ exceeds the threshold amount given by Eq. 20, however, the evaporation rate becomes so small that a crust no longer forms at the droplet surface. In this case, the droplet equilibrium radius rapidly increases with $RH/\gamma$, as expected in the absence of crust formation (see Eq. 3). Figure 7b indicates that an increase in the initial volume fraction of solutes increases the droplet equilibrium size, as expected.

To examine how the solute-concentration dependence of the internal water diffusivity affects the results, the equilibrium radius is calculated using the numerical method described in Appendix C for both $\beta > 0$ and $\beta < 0$, i.e., considering that $D_w^{l,sol}$ decreases or increases with the solute concentration (see Eq. 19). The results are shown by open and solid symbols in Fig. 7 and demonstrate that the solute-concentration dependence of the internal water diffusivity can slightly accelerate the crust formation when $\beta > 0$, or delay it when $\beta < 0$. Neglecting such dependence, however, does not make a significant error in the results.

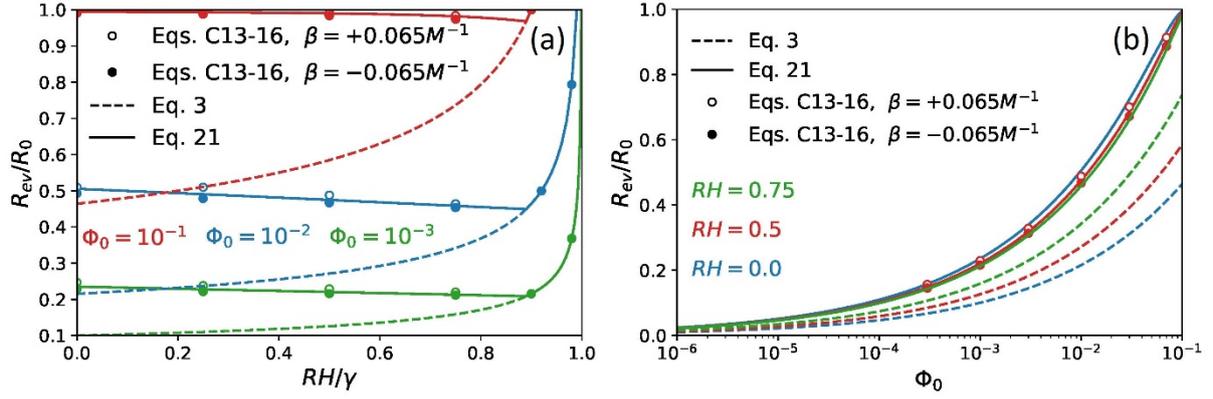

Figure 7 Variation of the droplet equilibrium radius $R_{ev}$ with (a) relative humidity $RH$ divided by the water activity coefficient $\gamma$ (b) initial volume fraction of solutes $\Phi_0$. Broken lines show the results from Eq. 3 that neglects the possibility of crust formation and solid lines show the results from Eq. 21, which is derived assuming that a solid crust forms on the droplet surface when the solute concentration at the surface reaches a critical value, taken here as the solubility limit of NaCl in water 6.1 M [46]. The results obtained from the numerical method described in Appendix C.2, which accounts for the solute-concentration dependence of the internal water-diffusion coefficient, are shown by circles. These results are calculated using $\beta = \pm 0.065 M^{-1}$ and $v_s = v_w = 3 \times 10^{-29} \, m^3$ (see Eq. 19). The value reported in table 2 is used for the water diffusion coefficient in pure water $D_w^l = D_w^{l,25°C} = 2.3 \times 10^{-9} \, m^2/s$.

The next step would be to model the evaporation process in the presence of a crust. Due to the diversity of possible crust structures and lack of experimental data on the crust properties, this step is difficult. As discussed in section 1, depending on the type of solutes, the crust formed during the drying process might be a dry crust, which forms due to crystallization of salts in a salt solution, or a gel-like wet skin, which is expected to form on the surface of polymer-containing droplets. In the case of dry salt crusts, the evaporation process would continue through capillary action [39, 44]. To model the evaporation in the presence of such a crust, one would need to know about the crust porosity, its characteristic pore size, and the amount of water vapor or liquid water located inside the pores. In the case of wet crusts, the evaporation continues through water diffusion in the gel phase, which is easier to model. However, one would still need to know the equilibrium free energy, variation of water activity coefficient with the local concentration, and the water diffusion constant within the crust. Regardless of the crust type, one also needs to account for cavity nucleation and growth in the wet core [47], which occurs when the crust is strong enough to withstand the pressure difference caused by the internal water evaporation, as assumed in the present study. In addition, to provide an efficient model for water evaporation after the crust formation, one would need to consider the other possible fates for the crust, such as crust collapse (Fig.1, scenario e) and wrinkling (Fig. 1, scenario f). Therefore, modelling of the water evaporation process after crust formation is left for future research.

## 3. Conclusion

We have studied the effects of non-volatile solutes on water evaporation from a solute-containing droplet by introducing a droplet evaporation model that accounts for evaporation cooling, internal water concentration gradients, solute-induced water vapor-pressure reduction, and the solute-concentration dependence of the water diffusivity inside the droplet. For this, the evaporating droplet is assumed to contain two regions: an outer shell where the liquid phase exhibits a concentration gradient and an internal core where the concentration profile remains uniform. Accordingly, the water evaporation is considered as a two-stage process. In the first stage, the outer shell grows from the surface towards the core of the droplet. When the shell covers almost the whole droplet, evaporation turns into the second stage, where the water concentration in the internal core gradually decreases to its equilibrium value.

Our investigations indicate that the presence of non-volatile solutes (such as salt, proteins, peptides, virions, etc) within a drying aqueous droplet has different, and sometimes contradictory, effects on the evaporation process. The most well-known effect is the water vapor-pressure reduction due to the dilution of the liquid water in the presence of solutes, which considerably slows down the evaporation process. This is, however, not the only effect. The solutes in the liquid phase are found to affect the evaporation-induced droplet cooling and decrease the temperature depression at the droplet surface, which results in an increased evaporation rate and, thus, slightly expedites the evaporation process. Our investigations indicate that this effect is not significant compared to the first effect, but still considerable.

Another important effect of solutes on the drying process is that in the presence of non-volatile solutes, the water concentration at the droplet surface deviates from the mean water concentration in the droplet due to the finite internal water diffusivity. This leads to a water concentration gradient in the liquid phase, which slows down the evaporation process. This effect can be intensified by any factor that decreases the internal water diffusivity, such as the presence of strongly hydrated solutes. The solute-concentration dependence of the water diffusion coefficient is, therefore, studied as a factor that influences the evaporation process. However, our results reveal that this effect is negligible and can be safely neglected. According to our investigations, in the presence of strongly hydrated solutes, the water-diffusion coefficient at the droplet surface decreases due to the increased solute surface concentration. On the other hand, the presence of solutes is found to decrease the thickness of the outer shell and, consequently, increases the water concentration gradient at the droplet surface. As a result, the water evaporation flux remains almost independent of the solute-induced variations of the local diffusion coefficient.

When the solute concentration at the droplet surface exceeds the solubility limit of solutes, a crust forms on the surface, which changes the evaporation mechanism. A crust affects the morphology of the dry particle formed at the end of the drying process by producing a hollow particle with a size larger than expected in the absence of crust. Our results reveal three factors that can delay the crust formation: a decrease in the initial volume fraction of solutes, an increase in the relative humidity, or a decrease in the water activity coefficient. Also, the crust formation is found to be slightly accelerated in the presence of solutes that decrease the internal water diffusivity, and slightly delayed in the presence of solutes that increase the internal water diffusivity.

Although the present study provides insight into the effects of non-volatile solutes on the evaporation process of solute-containing droplets, further work is required to answer a number of important open questions: (I) How does the water activity coefficient vary during the lifetime of the droplet and how do non-ideal effects due to solute-water interactions affect the drying process? (II) How does the evaporation process continue after the crust formation? What is the exact mechanism of water evaporation in the presence of a dry crust? Does it involve capillary action? If so, more information about the crust porosity, its characteristic pore size, and the amount of water located inside the pores is required. (III) What happens after the formation of a gel-like skin including saturated polymers?

**Acknowledgements**

We gratefully acknowledge funding by the Deutsche Forschungsgemeinschaft (DFG) via grant NE810/11 and by the ERC Advanced Grant NoMaMemo No. 835117.

**References**

[1] W. J. Massman, "A review of the molecular diffusivities of H2O, CO2, CH4, CO, O3, SO2, NH3, N2O, NO, and NO2 in air, O2 and N2 near STP," *Atmospheric Environment,* vol. 32, no. 6, pp. 1111-1127, 1998/03/01/ 1998, doi: https://doi.org/10.1016/S1352-2310(97)00391-9.

[2] R. A. Robinson and R. H. Stokes, *Electrolyte Solutions: Second Revised Edition*. Dover Publications, Incorporated, 2012.

[3] L. Glasser, "The effective volumes of waters of crystallization: non-ionic pharmaceutical systems," *Acta Crystallographica Section B,* vol. 75, no. 5, pp. 784-787, 2019, doi: doi:10.1107/S2052520619010436.

[4] L. C. Thomas, *Fundamentals of heat transfer*. United States: Prentice-Hall, Inc.,Englewood Cliffs, NJ (in English), 1980.

[5] D. R. Lide, *Handbook of Chemistry & Physics*, 89 ed. CRC, 2008.

[6] R. R. Netz, "Mechanisms of Airborne Infection via Evaporating and Sedimenting Droplets Produced by Speaking," *The Journal of Physical Chemistry B,* vol. 124, no. 33, pp. 7093-7101, 2020/08/20 2020, doi: 10.1021/acs.jpcb.0c05229.

[7] D. Mampallil, "Some physics inside drying droplets," *Resonance,* vol. 19, no. 2, pp. 123-134, 2014/02/01 2014, doi: 10.1007/s12045-014-0016-z.

[8] F. K. A. Gregson, J. F. Robinson, R. E. H. Miles, C. P. Royall, and J. P. Reid, "Drying Kinetics of Salt Solution Droplets: Water Evaporation Rates and Crystallization," *The Journal of Physical Chemistry B,* vol. 123, no. 1, pp. 266-276, 2019/01/10 2019, doi: 10.1021/acs.jpcb.8b09584.

[9] A. Gharsallaoui, G. Roudaut, O. Chambin, A. Voilley, and R. Saurel, "Applications of spray-drying in microencapsulation of food ingredients: An overview," *Food Research International,* vol. 40, no. 9, pp. 1107-1121, 2007/11/01/ 2007, doi: https://doi.org/10.1016/j.foodres.2007.07.004.

[10] R. P. Sear and P. B. Warren, "Diffusiophoresis in nonadsorbing polymer solutions: The Asakura-Oosawa model and stratification in drying films," *Physical Review E,* vol. 96, no. 6, p. 062602, 12/06/ 2017, doi: 10.1103/PhysRevE.96.062602.

[11] D. Zang, S. Tarafdar, Y. Y. Tarasevich, M. Dutta Choudhury, and T. Dutta, "Evaporation of a Droplet: From physics to applications," *Physics Reports,* vol. 804, pp. 1-56, 2019/04/29/ 2019, doi: https://doi.org/10.1016/j.physrep.2019.01.008.

[12] W. Liu, J. Shen, and S. R. Bhatia, "An in-situ SAXS approach to probe stratification during drying of inorganic nanoparticle films," *Inorganica Chimica Acta,* vol. 517, p. 120213, 2021/03/01/ 2021, doi: https://doi.org/10.1016/j.ica.2020.120213.

[13] R. R. Netz and W. A. Eaton, "Physics of virus transmission by speaking droplets," *Proceedings of the National Academy of Sciences,* vol. 117, no. 41, p. 25209, 2020, doi: 10.1073/pnas.2011889117.

[14] A. Božič and M. Kanduč, "Relative humidity in droplet and airborne transmission of disease," *Journal of Biological Physics,* vol. 47, no. 1, pp. 1-29, 2021/03/01 2021, doi: 10.1007/s10867-020-09562-5.

[15] V. Stadnytskyi, C. E. Bax, A. Bax, and P. Anfinrud, "The airborne lifetime of small speech droplets and their potential importance in SARS-CoV-2 transmission," *Proceedings of the National Academy of Sciences,* vol. 117, no. 22, p. 11875, 2020, doi: 10.1073/pnas.2006874117.

[16] P. Anfinrud, V. Stadnytskyi, C. E. Bax, and A. Bax, "Visualizing Speech-Generated Oral Fluid Droplets with Laser Light Scattering," *New England Journal of Medicine,* vol. 382, no. 21, pp. 2061-2063, 2020/05/21 2020, doi: 10.1056/NEJMc2007800.

[17] L. Liu, J. Wei, Y. Li, and A. Ooi, "Evaporation and dispersion of respiratory droplets from coughing," *Indoor Air,* https://doi.org/10.1111/ina.12297 vol. 27, no. 1, pp. 179-190, 2017/01/01 2017, doi: https://doi.org/10.1111/ina.12297.

[18] R. D. Deegan, O. Bakajin, T. F. Dupont, G. Huber, S. R. Nagel, and T. A. Witten, "Capillary flow as the cause of ring stains from dried liquid drops," *Nature,* vol. 389, no. 6653, pp. 827-829, 1997/10/01 1997, doi: 10.1038/39827.

[19] H. Y. Erbil, "Evaporation of pure liquid sessile and spherical suspended drops: A review," *Advances in Colloid and Interface Science,* vol. 170, no. 1, pp. 67-86, 2012/01/15/ 2012, doi: https://doi.org/10.1016/j.cis.2011.12.006.

[20] S. Tarafdar, Y. Y. Tarasevich, M. Dutta Choudhury, T. Dutta, and D. Zang, "Droplet Drying Patterns on Solid Substrates: From Hydrophilic to Superhydrophobic Contact to Levitating Drops," *Advances in Condensed Matter Physics,* vol. 2018, p. 5214924, 2018/04/03 2018, doi: 10.1155/2018/5214924.

[21] M. M. Parsa, S. Harmand, and K. Sefiane, "Mechanisms of pattern formation from dried sessile drops," *Advances in Colloid and Interface Science,* vol. 254, pp. 22-47, 03/24 2018, doi: 10.1016/j.cis.2018.03.007.

[22] T. Squires, "Drops on soft surfaces learn the hard way," *Proceedings of the National Academy of Sciences of the United States of America,* vol. 110, 07/19 2013, doi: 10.1073/pnas.1310672110.

[23] D. Quéré, "Wetting and Roughness," *Annual Review of Materials Research,* vol. 38, no. 1, pp. 71-99, 2008/08/01 2008, doi: 10.1146/annurev.matsci.38.060407.132434.

[24] S. Karpitschka, F. Liebig, and H. Riegler, "Marangoni Contraction of Evaporating Sessile Droplets of Binary Mixtures," *Langmuir,* vol. 33, pp. 4682-4687, 2017. [Online]. Available: https://publishup.uni-potsdam.de/frontdoor/index/index/docId/46656.

[25] X. Chen, X. Wang, P. G. Chen, and Q. Liu, "Thermal effects of substrate on Marangoni flow in droplet evaporation: Response surface and sensitivity analysis," *International Journal of Heat and Mass Transfer,* vol. 113, pp. 354-365, 2017/10/01/ 2017, doi: https://doi.org/10.1016/j.ijheatmasstransfer.2017.05.076.

[26] Y. Wei, W. Deng, and R.-H. Chen, "Effects of internal circulation and particle mobility during nanofluid droplet evaporation," *International Journal of Heat and Mass Transfer,* vol. 103, pp. 1335-1347, 2016/12/01/ 2016, doi: https://doi.org/10.1016/j.ijheatmasstransfer.2016.08.037.

[27] U. Maurice, M. Mezhericher, A. Levy, and I. Borde, "Drying of Droplets Containing Insoluble Nanoscale Particles: Second Drying Stage," *Drying Technology,* vol. 33, no. 15-16, pp. 1837-1848, 2015/11/18 2015, doi: 10.1080/07373937.2015.1039540.

[28]  K. Roger, E. Sparr, and H. Wennerström, "Evaporation, diffusion and self-assembly at drying interfaces," *Physical Chemistry Chemical Physics,* 10.1039/C8CP00305J vol. 20, no. 15, pp. 10430-10438, 2018, doi: 10.1039/C8CP00305J.

[29]  S. Mahiuddin and K. Ismail, "Concentration dependence of viscosity of aqueous electrolytes. A probe into the higher concentration," *The Journal of Physical Chemistry,* vol. 87, no. 25, pp. 5241-5244, 1983/12/01 1983, doi: 10.1021/j150643a036.

[30]  J. S. Kim, Z. Wu, A. R. Morrow, A. Yethiraj, and A. Yethiraj, "Self-Diffusion and Viscosity in Electrolyte Solutions," *The Journal of Physical Chemistry B,* vol. 116, no. 39, pp. 12007-12013, 2012/10/04 2012, doi: 10.1021/jp306847t.

[31]  O. Miyawaki, A. Saito, T. Matsuo, and K. Nakamura, "Activity and Activity Coefficient of Water in Aqueous Solutions and Their Relationships with Solution Structure Parameters," *Bioscience, Biotechnology, and Biochemistry,* vol. 61, no. 3, pp. 466-469, 1997/01/01 1997, doi: 10.1271/bbb.61.466.

[32]  G. Brenn, "Concentration fields in evaporating droplets," *International Journal of Heat and Mass Transfer - INT J HEAT MASS TRANSFER,* vol. 48, pp. 395-402, 01/31 2005, doi: 10.1016/j.ijheatmasstransfer.2004.07.039.

[33]  X. Jiang, T. Ward, F. Swol, and C. Brinker, "Numerical Simulation of Ethanol–Water–NaCl Droplet Evaporation," *Industrial & Engineering Chemistry Research - IND ENG CHEM RES,* vol. 49, 05/19 2010, doi: 10.1021/ie902042z.

[34]  Y. Wei, W. Deng, and R.-H. Chen, "Effects of insoluble nano-particles on nanofluid droplet evaporation," *International Journal of Heat and Mass Transfer,* vol. 97, pp. 725-734, 2016/06/01/ 2016, doi: https://doi.org/10.1016/j.ijheatmasstransfer.2016.02.052.

[35]  G. Boulet, A. Chehbouni, I. Braud, B. Duchemin, and A. Lakhal, "Evaluation of a two-stage evaporation approximation for contrasting vegetation cover," *Water Resources Research,* https://doi.org/10.1029/2004WR003212 vol. 40, no. 12, 2004/12/01 2004, doi: https://doi.org/10.1029/2004WR003212.

[36]  C. J. Homer, X. Jiang, T. L. Ward, C. J. Brinker, and J. P. Reid, "Measurements and simulations of the near-surface composition of evaporating ethanol–water droplets," *Physical Chemistry Chemical Physics,* 10.1039/B904070F vol. 11, no. 36, pp. 7780-7791, 2009, doi: 10.1039/B904070F.

[37]  A. Alowaisy and N. Yasufuku, "Characteristics of the second stage of evaporation and water redistribution through double layered sandy soil profiles," *Lowland Technology International,* vol. 20, pp. 273-284, 01/01 2018.

[38]  E. P. Vejerano and L. C. Marr, "Physico-chemical characteristics of evaporating respiratory fluid droplets," *Journal of The Royal Society Interface,* vol. 15, no. 139, p. 20170939, 2018/02/28 2018, doi: 10.1098/rsif.2017.0939.

[39]  E. Boel *et al.*, "Unraveling Particle Formation: From Single Droplet Drying to Spray Drying and Electrospraying," *Pharmaceutics,* vol. 12, no. 7, 2020, doi: 10.3390/pharmaceutics12070625.

[40]  D. Sen, S. Mazumder, J. S. Melo, A. Khan, S. Bhattyacharya, and S. F. D'Souza, "Evaporation Driven Self-Assembly of a Colloidal Dispersion during Spray Drying: Volume Fraction Dependent Morphological Transition," *Langmuir,* vol. 25, no. 12, pp. 6690-6695, 2009/06/16 2009, doi: 10.1021/la900160z.

[41]  N. Shahidzadeh, M. F. L. Schut, J. Desarnaud, M. Prat, and D. Bonn, "Salt stains from evaporating droplets," *Scientific Reports,* vol. 5, no. 1, p. 10335, 2015/05/27 2015, doi: 10.1038/srep10335.

[42]  T. Okuzono, K. y. Ozawa, and M. Doi, "Simple Model of Skin Formation Caused by Solvent Evaporation in Polymer Solutions," *Physical Review Letters,* vol. 97, no. 13, p. 136103, 09/26/ 2006, doi: 10.1103/PhysRevLett.97.136103.

[43]  Y. Zhang, X. Gao, H. Chu, and B. P. Binks, "Various crust morphologies of colloidal droplets dried on a super-hydrophobic surface," *Canadian Journal of Physics,* vol. 98, no. 11, pp. 1055-1059, 2020/11/01 2020, doi: 10.1139/cjp-2019-0451.

[44] M. Mezhericher, A. Levy, and I. Borde, "Theoretical Models of Single Droplet Drying Kinetics: A Review," *Drying Technology,* vol. 28, no. 2, pp. 278-293, 2010/03/10 2010, doi: 10.1080/07373930903530337.

[45] S. P. Humphrey and R. T. Williamson, "A review of saliva: Normal composition, flow, and function," *The Journal of Prosthetic Dentistry,* vol. 85, no. 2, pp. 162-169, 2001/02/01/ 2001, doi: https://doi.org/10.1067/mpr.2001.113778.

[46] M. Bergstad, D. Or, P. Withers, and N. Shokri, "The influence of NaCl concentration on salt precipitation in heterogeneous porous media," *Water Resources Research,* vol. 53, 02/01 2017, doi: 10.1002/2016WR020060.

[47] F. Meng, M. Doi, and Z. Ouyang, "Cavitation in Drying Droplets of Soft Matter Solutions," *Physical review letters,* vol. 113, p. 098301, 08/29 2014, doi: 10.1103/PhysRevLett.113.098301.

[48] M. Mulansky and K. Ahnert, "Odeint library," in *Scholarpedia* vol. 9, ed, 2014.

[49] P. C. Ellgen, *Thermodynamics and Chemical Equilibrium*. Createspace Independent Pub, 2014.

**Appendix A: Water evaporation from a solute-containing droplet with sufficiently rapid internal diffusivity**

In this section, it is assumed that the water diffusion inside the droplet is sufficiently rapid, so that the water concentration inside the droplet is homogeneous. Also, we use the diffusion-limited approximation defined by $k_c R > D_w$, with $k_c$ being the condensation reaction rate coefficient, $R$ being the droplet radius, and $D_w$ being the molecular water diffusion constant in air. In this case, the total water evaporation flux from a solute-containing droplet can be expressed as [6]

$$J = 4\pi R D_w \left( c_g \left( 1 - \frac{\Phi_0 R_0^3}{R^3} \right) - c_0 \right) \tag{A1}$$

where $\Phi_0$ is the initial volume fraction of solutes, $R_0$ is the initial droplet radius, $c_0$ is the ambient water vapor concentration, and $c_g$ is the saturated water vapor concentration in the absence of solutes. To account for non-ideal effects caused by water-solute interactions, the term $1 - \Phi_0 R_0^3/R^3$ in equation A1, which reflects the solute-induced water vapor-pressure reduction, must be multiplied by the activity coefficient of water $\gamma$ in the presence of solutes, as will be discussed in section B.1. Another factor that should be accounted for in the calculations is the evaporation-induced temperature reduction at the droplet surface. The approach proposed in reference [6] is employed here to account for this factor, as follows.

The temperature distribution around a spherical heat sink (here, an evaporating droplet) can be readily obtained from the heat diffusion equation

$$\frac{\lambda}{r^2} \frac{d}{dr} r^2 \frac{d}{dr} T(r,t) = c C_P \frac{d}{dt} T(r,t) \tag{A2}$$

with $\lambda$ and $C_P$ being the heat conductivity and the heat capacity of the medium, respectively. The stationary temperature distribution around the droplet can be, therefore, obtained from the steady-state form of equation A2, resulting in

$$T(r) = T_0\left(1 - \frac{b_T}{r}\right) \tag{A3}$$

where $T_0$ is the ambient temperature and $b_T$ is an unknown constant to be determined. It is worth noting that here, we ignore the temperature gradient inside the droplet and thus the heat diffusion equation is solved only outside the droplet. Using the adiabatic approximation (i.e., assuming that the time it takes for the stationary temperature distribution to build up is negligible), the droplet surface temperature can be written according to Eq. A3 as

$$T_s = T_0\left(1 - \frac{b_T}{R}\right) \tag{A4}$$

From Eq. A4, one obtains the temperature depression at the droplet surface as

$$\Delta T = T_0 - T_s = \frac{T_0 b_T}{R} \tag{A5}$$

As discussed in reference [6], such temperature depression causes a reduction in the surface water vapor concentration $c_g^{surf}$, which can be expressed in terms of $\Delta T$ as

$$c_g^{surf} = c_g(1 - \varepsilon_c \Delta T) \tag{A6}$$

where $\varepsilon_c = 0.032\, K^{-1}$ is a numerical perfector obtained by linear interpolation of the water vapor densities at 0°C and at 25°C (see table 2 of the main text). The reduced temperature at the droplet surface produces a heat flux into the droplet, which is given by

$$J_h = 4\pi R^2 \lambda_{air} \frac{d}{dR} T(R) = 4\pi \lambda_{air} T_0 b_T = 4\pi \lambda_{air} R \Delta T \tag{A7}$$

with $\lambda_{air}$ being the heat conductivity of air. One can relate the heat flux into the droplet $J_h$ to the evaporation flux from the droplet surface $J$ using the energy balance in the stationary state, which reads

$$J_h = h_{ev} J \tag{A8}$$

with $h_{ev}$ being the molecular evaporation enthalpy of water. Replacing $c_g$ by $c_g^{surf}$ (given by Eq. A6) in Eq. A1 and combining the result with equations A7 and A8, the temperature depression at the droplet surface is obtained as

$$\Delta T = \gamma \varepsilon_T \left( \frac{1 - \frac{\Phi_0 R_0^3}{R^3} - \frac{RH}{\gamma}}{1 + \gamma \varepsilon_c \varepsilon_T \left(1 - \frac{\Phi_0 R_0^3}{R^3}\right)} \right) \tag{A9}$$

Here, $RH = c_0/c_g$ is the relative fractional air humidity and $\varepsilon_T$ is a numerical prefactor given by

$$\varepsilon_T \equiv \frac{D_w c_g h_{ev}}{\lambda_{air}} \tag{A10}$$

Using the values reported for $D_w$, $c_g$, $h_{ev}$, and $\lambda_{air}$ at 25 °C (see table 2 of the main text), one obtains $\varepsilon_T = 54\ K$. By incorporating $\Delta T$ from Eq. A10 into Eq. A1, the total evaporation flux can be rewritten as

$$J = 4\pi D_w c_g \gamma R \left( \frac{1 - \frac{\Phi_0 R_0^3}{R^3} - \frac{RH}{\gamma}}{1 + \gamma \varepsilon_c \varepsilon_T \left(1 - \frac{\Phi_0 R_0^3}{R^3}\right)} \right) \quad (A11)$$

Comparing Eq. A11 to Eq. A1, one identifies the correction factor accounting for evaporation cooling as $\frac{1}{1+\gamma\varepsilon_c\varepsilon_T\left(1-\frac{\Phi_0 R_0^3}{R^3}\right)}$, which is dependent on the momentary volume fraction of solutes $\frac{\Phi_0 R_0^3}{R^3}$.

The next step is to calculate the evaporation time, i.e., the time it takes for the droplet radius to reach the equilibrium droplet radius. For this, we write the total mass conservation of the droplet in terms of the droplet volume

$$\frac{d}{dt}\left(\frac{4\pi R^3(t)}{3}\right) = -J v_w \quad (A12)$$

where $v_w$ is the volume of a water molecule in the liquid phase. Incorporating Eq. A11 into Eq. A12, one obtains the differential equation

$$\frac{dR}{dt} = -\frac{D_w c_g v_w \gamma \left(1 - \frac{RH}{\gamma}\right)\left(1 - \frac{R_{ev}^3}{R^3}\right)}{R\left(1 + \gamma\varepsilon_c\varepsilon_T\left(1 - \frac{\Phi_0 R_0^3}{R^3}\right)\right)} \quad (A13)$$

where $R_{ev}$ is the droplet equilibrium radius, meaning the droplet radius at which the evaporation flux (given by Eq. A11) vanishes

$$R_{ev} = R_0 \left(\frac{\Phi_0}{1 - \frac{RH}{\gamma}}\right)^{1/3} \quad (A14)$$

The differential equation A13 is not analytically solvable. Instead, we solve this equation numerically using the Python library "odeint" [48].

Neglecting the solute-concentration dependence of evaporation cooling, meaning replacing the correction factor $\frac{1}{1+\gamma\varepsilon_c\varepsilon_T\left(1-\frac{\Phi_0 R_0^3}{R^3}\right)}$ in equation A11 by that in the absence of solutes $\frac{1}{1+\gamma\varepsilon_c\varepsilon_T}$, one can obtain the analytically solvable differential equation

$$\frac{2RdR}{1-\frac{R_{ev}^3}{R^3}} = -\theta\left(1-\frac{RH}{\gamma}\right)dt \tag{A15}$$

where $\theta$ is a numerical prefactor given by

$$\theta = 2\gamma D_w c_g v_w \left(\frac{1}{1+\gamma\varepsilon_C\varepsilon_T}\right) \tag{A16}$$

The solution of the differential equation A15 can be written as [6]

$$t(R) = \frac{R_{ev}^2}{\theta\left(1-\frac{RH}{\gamma}\right)}\left[\mathcal{L}\left(\frac{R_0}{R_{ev}}\right) - \mathcal{L}\left(\frac{R}{R_{ev}}\right)\right] \tag{A17}$$

with $\mathcal{L}$ being a scaling function given by

$$\mathcal{L}(x) = x^2 - \frac{2}{\sqrt{3}}\arctan\left(\frac{1+\frac{2}{x}}{\sqrt{3}}\right) - \frac{1}{3}\ln\left(\frac{x^2+x+1}{(x-1)^2}\right) \tag{A18}$$

As discussed in reference [6], $\mathcal{L}(x)$ can be approximated as

$$\mathcal{L}(x) \cong x^2 + \frac{2}{3}\ln(1 - 1/x) \tag{A19}$$

The radius-dependent evaporation time of the droplet can be, therefore, estimated from Eq. A17 as

$$t(R) = \frac{R_0^2}{\theta\left(1-\frac{RH}{\gamma}\right)}\left[1 - \left(\frac{R}{R_0}\right)^2 - \frac{2}{3}\left(\frac{R_{ev}}{R_0}\right)^2 \ln\left(\frac{R_0(R-R_{ev})}{R(R_0-R_{ev})}\right)\right] \tag{A20}$$

As mentioned before, equation A20 is derived neglecting the effect of solutes on the evaporation cooling, which is an approximation that is discussed in section 2 of the main text. To account for this effect, one should replace $\theta$ in Eq. A15 with $\theta^{sol}$ given by Eq. A21, as follows from Eq. A13.

$$\theta^{sol} = \frac{2\gamma D_w c_g v_w}{1+\gamma\varepsilon_C\varepsilon_T\left(1-\frac{\Phi_0 R_0^3}{R^3}\right)} \tag{A21}$$

**Appendix B: Modeling of evaporation-induced concentration gradients inside a solute-containing droplet**

The main assumption in section A is that the internal mixing due to diffusion inside the droplet is sufficiently fast, so that the solute particles remain evenly distributed throughout the liquid phase. However, a fast water evaporation from the droplet surface will increase the solute concentration at the surface, which causes a concentration gradient within the droplet. In this section, the model used to account for such concentration gradient is introduced.

**B.1. The diffusion equation**

To model the droplet evaporation, the water concentration profiles inside and outside the droplet must be calculated. For this, one needs to solve the diffusion equation in radial coordinates

$$\frac{dc(r,t)}{dt} = \frac{1}{r^2}\frac{\partial}{\partial r}\left(D[c(r,t)]r^2\frac{\partial c(r,t)}{\partial r}\right) \tag{B1}$$

where $c(r,t)$ is the concentration profile and $D$ is the molecular particle diffusion coefficient, which might be concentration-dependent. In this section, the water diffusion coefficients in both air and liquid phase are assumed constant (the concentration dependence of the water diffusion coefficient in the liquid phase is discussed in section C). The stationary water vapor concentration profile $c_v(r)$ around a spherical droplet can be readily determined from the steady-state form of Eq. B1, resulting in

$$c_v(r) = c_0\left(1 + \frac{b_1}{r}\right) \tag{B2}$$

where $c_0$ is the ambient vapor concentration and $b_1$ is an unknown parameter to be determined. The same approach is used to calculate the water concentration profile inside the droplet. However, equation B1 has a singularity at $r = 0$, at the center of the droplet, which is removed by dividing the liquid phase into two regions: an internal core, where the concentration profile is uniform, and an outer shell, where the water concentration is dependent on the radial distance $r$ (see Fig. 4 of the main text). By solving the steady-state form of the diffusion equation in the outer shell, the water concentration profile inside the droplet is obtained as

$$c_l(r) = \begin{cases} c_i & r < R_i \\ b_2\left(1 + \frac{b_3}{r}\right) & R_i < r < R \end{cases} \tag{B3}$$

with $c_i$ and $R_i$ being the internal water concentration and the radius of the internal core, respectively, $R$ being the droplet radius, and $b_2$ and $b_3$ being unknown parameters to be determined. Equation B3 must satisfy the boundary conditions at $r = R$ and $r = R_i$

$$j(R^-) = j(R^+) \Rightarrow -D_w^l\left.\frac{\partial c_l}{\partial r}\right|_{r=R} = -D_w\left.\frac{\partial c_v}{\partial r}\right|_{r=R} \tag{B4}$$

$$c_l(R_i^-) = c_l(R_i^+) \tag{B5}$$

where $j$ is the flux density, and $D_w^l$ and $D_w$ are the water diffusion coefficients in the liquid droplet and in air, respectively. Using the above boundary conditions, the liquid water concentration profile can be rewritten as

$$c_l(r) = \begin{cases} c_i & r < R_i \\ c_i + \dfrac{D_w}{D_w^l} c_0 b_1 \left(\dfrac{1}{r} - \dfrac{1}{R_i}\right) & R_i < r < R \end{cases} \tag{B6}$$

The remaining unknown parameter $b_1$ can be determined from the reactive boundary condition at the droplet surface

$$j = -D_w \left.\dfrac{\partial c_v}{\partial r}\right|_{r=R} = k_e c_l(R) - k_c c_v(R) \tag{B7}$$

with $k_e$ and $k_c$ being the evaporation and the condensation reaction rate coefficients, respectively. Inserting $c_v(R)$ and $c_l(R)$ from the concentration profiles B2 and B6 into equation B7, one obtains $b_1$ as

$$b_1 = \dfrac{R^2}{c_0} \left( \dfrac{k_e c_i - k_c c_0}{D_w + \left(k_c - k_e \dfrac{D_w}{D_w^l}\left(1 - \dfrac{R}{R_i}\right)\right) R} \right) \tag{B8}$$

Another condition that must be satisfied is that in the evaporation equilibrium state, meaning in the presence of saturated water vapor, the evaporation flux must vanish. Here, the saturated water vapor concentration in the presence of solutes $c_g^{sol}$ is calculated using the formulation provided in reference [6] instead of the commonly used Raoult's law [49]. Accordingly, the water chemical potential $\mu_w^l$ in a two-component liquid system follows as

$$\dfrac{\mu_w^l}{k_B T} = -\dfrac{\partial S}{k_B \partial N_w} + \dfrac{\mu_{ex}}{k_B T} = \ln\left(\dfrac{1-\Phi}{v_w}\right) + \Phi\left(1 - \dfrac{v_w}{v_s}\right) + \dfrac{\mu_{ex}}{k_B T} \tag{B9}$$

where $S$ is the entropy of the system, $\Phi$ is the volume fraction of solutes, $v_s$ is the molecular volume of solutes, $\mu_{ex}$ is the water excess chemical potential, and $k_B T$ is the thermal energy. For $v_s \geq 0.1 v_w$, the term $\Phi\left(1 - \dfrac{v_w}{v_s}\right)$ in Eq. B9 is negligible compared to the term $\ln\left(\dfrac{1-\Phi}{v_w}\right)$. So, one can instead rewrite this equation as

$$\dfrac{\mu_w^l}{k_B T} \approx \ln\left(\dfrac{1-\Phi}{v_w}\right) + \dfrac{\mu_{ex}}{k_B T} \tag{B10}$$

From the ideal expression for the water vapor chemical potential

$$\dfrac{\mu_w^g}{k_B T} = \ln(c_g) \tag{B11}$$

and the equality of chemical potentials, $\mu_w^g = \mu_w^l$, the equilibrium vapor concentration in the presence of solutes follows from equations B10 and B11 as

$$c_g^{sol} = \left(\dfrac{1-\Phi}{v_w}\right) e^{\mu_{ex}/k_B T} \tag{B12}$$

Equation B12 is derived assuming ideal mixing and ideal volume additivity. In the case of non-ideal mixtures, one can use the same relation by introducing the so-called activity coefficient γ, which accounts for the inter-component interactions (here, the interactions between water and solutes)

$$c_g^{sol} = \gamma \left(\frac{1-\Phi}{v_w}\right) e^{\mu_{ex}/k_B T} \tag{B13}$$

Therefore, in the evaporation-equilibrium state one can write

$$c_g^{sol} = \gamma(1-\Phi^{ev})c_g \tag{B14}$$

$$c_l^{ev} = \frac{1}{v_w}(1-\Phi^{ev}) \tag{B15}$$

where $\Phi^{ev} = \Phi_0 \frac{R_0^3}{R_{ev}^3}$ and $c_l^{ev}$ are the volume fraction of solutes and the solute concentration at the equilibrium state, respectively, and $c_g = \frac{e^{\mu_{ex}/k_B T}}{v_w}$ represents the water vapor concentration in the absence of solutes. In the case of multi-component droplets, $\Phi^{ev}$ in equations B14 and B15 is replaced by $\sum_i \Phi_i^{ev}$, with $i$ denoting the solute components.

Having $c_g^{sol}$ and $c_l^{ev}$ from equations B14 and B15, one can relate $k_e$ and $k_c$ by equating the evaporation rate with the condensation rate in the equilibrium state $k_e c_l^{ev} = k_c c_g^{sol}$. As a result, $k_e$ can be expressed in terms of $k_c$ as

$$k_e = \gamma v_w k_c c_g \tag{B16}$$

Equation B16 shows that the evaporation reaction rate coefficient in the presence of solutes equals its value for a pure water droplet $v_w k_c c_g$ times the water activity coefficient γ. An important point here is that the water activity coefficient may decrease or increase during the lifetime of the droplet. Indeed, an increase in the concentration of strongly hydrated solutes decreases γ due to the reduced volatility of water molecules. The opposite is true for weakly hydrated solutes because such solutes can disrupt hydrogen bonds. Therefore, the evaporation-induced increase in the solute concentration may change the activity coefficient of water and, consequently, affect the water evaporation rate. To achieve an accurate model of evaporation, therefore, one needs to also account for the concentration-dependent variation of the water activity coefficient during the lifetime of the drying droplet. In the following calculations, however, we consider a constant activity coefficient for the sake of simplicity.

Using Eq. B16, $k_e$ can be eliminated from Eq. B8, resulting in

$$b_1 = \frac{R^2 \gamma k_c c_g}{c_0} \left( \frac{v_w c_i - \frac{RH}{\gamma}}{D_w + \left(1 - \frac{\gamma}{\alpha}\left(1 - \frac{R}{R_i}\right)\right) k_c R} \right) \tag{B17}$$

where the numerical prefactor $\alpha$ is given by

$$\alpha = \frac{D_w^l}{D_w c_g v_w} \tag{B18}$$

Considering the values of $D_w^l$, $D_w$, $c_g$, and $v_w$ at 25 °C, which are listed in table 2 of the main text, $\alpha$ for pure water at 25 °C is approximately $\alpha \approx 4$. However, in the present study, we use different values of $D_w^l$ and, consequently, different values of $\alpha$, to investigate the effect of the internal diffusion constant on the water evaporation process.

Having $b_1$ from Eq. B17, one can readily obtain the total water evaporation flux as

$$J = 4\pi R^2 \times -D_w \left.\frac{\partial c_v}{\partial r}\right|_{r=R} = 4\pi D_w \gamma k_c c_g R^2 \left(\frac{v_w c_i - \frac{RH}{\gamma}}{D_w + \left(1 - \frac{\gamma}{\alpha}\left(1 - \frac{R}{R_i}\right)\right)k_c R}\right) \tag{B19}$$

**B.2 The diffusion-limited rate scenario**

As mentioned in section A, here we focus on the diffusion-limited regime defined by $D_w < k_c R$. Since the radius of the internal core $R_i$ is equal to or smaller than the droplet radius $R$, the diffusion-limited approximation results in

$$D_w < k_c R < \left(1 - \frac{\gamma}{\alpha}\left(1 - \frac{R}{R_i}\right)\right)k_c R \tag{B20}$$

Therefore, $D_w$ in the denominator of equation B19 can be safely neglected and after simplifications, this equation yields

$$J = \frac{4\pi D_w^l R}{v_w}\left(\frac{v_w c_i - \frac{RH}{\gamma}}{\frac{\alpha}{\gamma} - 1 + \frac{R}{R_i}}\right) \tag{B21}$$

After applying the diffusion-limited approximation to Eq. B17 and incorporating the result into Eq. B6, the water concentration profile in the liquid phase can be rewritten as

$$c_l(r) = \begin{cases} c_i & r < R_i \\ c_i + \dfrac{\left(v_w c_i - \frac{RH}{\gamma}\right)\left(\frac{R}{r} - \frac{R}{R_i}\right)}{v_w\left(\frac{\alpha}{\gamma} - 1 + \frac{R}{R_i}\right)} & R_i < r < R \end{cases} \tag{B22}$$

**B.3 Evaporation cooling effects**

The calculations provided in sections B.1 and B.2 were done without considering the effect of evaporation cooling. To account for this effect, the same approach as that explained in section

A is used here. Replacing $c_g$ by $c_g^{surf}$ defined by Eq. A6 and writing the energy balance in the stationary state (Eq. A8), the temperature reduction at the droplet surface is obtained as

$$\Delta T = \frac{\alpha \varepsilon_T (v_w c_i - \frac{RH}{\gamma})}{\frac{\alpha}{\gamma} - 1 + \frac{R}{R_i} + \alpha \varepsilon_T \varepsilon_c v_w c_i} \tag{B23}$$

After inserting $\Delta T$ from Eq. B23 into Eq. A6 and incorporating the result into Eq. B21, the total evaporation flux in the presence of evaporation-induced droplet cooling can be written as

$$J = \frac{4\pi D_w^l R}{v_w} \left( \frac{v_w c_i - \frac{RH}{\gamma}}{\frac{\alpha}{\gamma} - 1 + \frac{R}{R_i} + \alpha \varepsilon_T \varepsilon_c v_w c_i} \right) \tag{B24}$$

The term that accounts for evaporation cooling is thus given by $\alpha \varepsilon_T \varepsilon_c v_w c_i$, which appears in the dominator of Eq. B24 and reduces the water evaporation flux. By including this term in the water concentration profile (Eq. B22), one obtains

$$c_l(r) = \begin{cases} c_i & r < R_i \\ c_i + \dfrac{\left(v_w c_i - \frac{RH}{\gamma}\right)\left(\frac{R}{r} - \frac{R}{R_i}\right)}{v_w \left(\frac{\alpha}{\gamma} - 1 + \frac{R}{R_i} + \alpha \varepsilon_T \varepsilon_c v_w c_i\right)} & R_i < r < R \end{cases} \tag{B25}$$

### B.4. Mass conservation

To complete our calculations, we need to determine the time-dependent parameters that appear in the above equations. This will be done using the mass conservation equations for both solutes and water. Relevant calculations for the two drying stages introduced in section 2 of the main text are described below.

### B.4.1 The first drying stage

As stated in section 2 of the main text, the first drying stage models the shrinking of the droplet core. In this stage, we have two time-dependent parameters: the droplet radius $R$ and the radius of the internal core $R_i$. The water concentration in the internal core $c_i$ is fixed to the initial concentration, which is given by

$$c_i = \frac{1 - \Phi_0}{v_w} \tag{B26}$$

Based on equations B24 and B25, the total evaporation flux and the water concentration profile in the first drying stage can be expressed as

$$J^{1st}(t) = \frac{4\pi D_w^l R(t)}{v_w} \left( \frac{1 - \Phi_0 - \frac{RH}{\gamma}}{\frac{\alpha}{\gamma} - 1 + \frac{R(t)}{R_i(t)} + \alpha \varepsilon_T \varepsilon_c (1 - \Phi_0)} \right) \tag{B27}$$

$$c_l^{1st}(r,t) = \begin{cases} \dfrac{1-\Phi_0}{v_w} & r \leq R_i(t) \\[1em] \dfrac{1-\Phi_0}{v_w} + \dfrac{\left(1-\Phi_0-\dfrac{RH}{\gamma}\right)\left(\dfrac{R(t)}{r}-\dfrac{R(t)}{R_i(t)}\right)}{v_w\left(\dfrac{\alpha}{\gamma}-1+\dfrac{R(t)}{R_i(t)}+\alpha\varepsilon_T\varepsilon_c(1-\Phi_0)\right)} & R_i(t) < r \leq R(t) \end{cases} \qquad (B28)$$

The time-dependent parameters $R(t)$ and $R_i(t)$ are calculated in such a way that mass conservation for both solutes and water is satisfied. We assume that the droplet only includes non-volatile solutes. Therefore, mass conservation of the solutes at an arbitrary time reads

$$\int_0^R c_s(r) 4\pi r^2 dr = \frac{4\pi R_0^3 \Phi_0}{3 v_s} \Rightarrow \int_0^R (1 - v_w c_l(r)) r^2 dr = \frac{R_0^3 \Phi_0}{3} \qquad (B29)$$

with $c_s(r) = \frac{1}{v_s}(1 - v_w c_l(r))$ being the solute concentration profile. Note that Eq. B29 holds in both first and second drying stages. After calculating $c_s^{1st}(r)$ from Eq. B28 and inserting the result into Eq. B29, one obtains

$$\left(\frac{R_i}{R}\right)^3 + \left(f(R)\left(\frac{\alpha}{\gamma} - 1 + \alpha\varepsilon_T\varepsilon_c(1-\Phi_0)\right) - 3\right)\left(\frac{R_i}{R}\right) + f(R) + 2 = 0 \qquad (B30)$$

with $f(R)$ being a function of the droplet radius that is given by

$$f(R) = \frac{2\Phi_0\left(1 - \dfrac{R_0^3}{R^3}\right)}{1 - \Phi_0 - \dfrac{RH}{\gamma}} \qquad (B31)$$

The cubic equation B30 can be solved to yield $R_i$ as a function of $R$. The next step is to obtain $R$ as a function of $t$ from the total mass conservation equation for the droplet (Eq. A12), which gives rise to the differential equation

$$\frac{dR}{dt} = -\frac{D_w^l\left(1 - \Phi_0 - \dfrac{RH}{\gamma}\right)}{R\left(\dfrac{\alpha}{\gamma} - 1 + \dfrac{R}{R_i} + \alpha\varepsilon_T\varepsilon_c(1-\Phi_0)\right)} \qquad (B32)$$

Equation B32 is, however, not analytically solvable. Instead, we solve this equation numerically. For this, $R_i/R$ is first determined from Eq. B30 at different values of $R/R_0$. Then, a suitable polynomial is fitted to the resulting data set (see figure B1). By incorporating this function into Eq. B32, this equation is then numerically solved using the Python library "odeint" to calculate $R$ as a function of $t$. Finally, the time of the first drying stage is estimated as $\tau^{1st} = t(R°)$, with $R°$ being the droplet radius at the end of the first drying stage, i.e., when the radius of the internal core $R_i$ reaches the value $R_i°$ (see Fig. 4b of the main text), which can be readily calculated from Eq. B30.

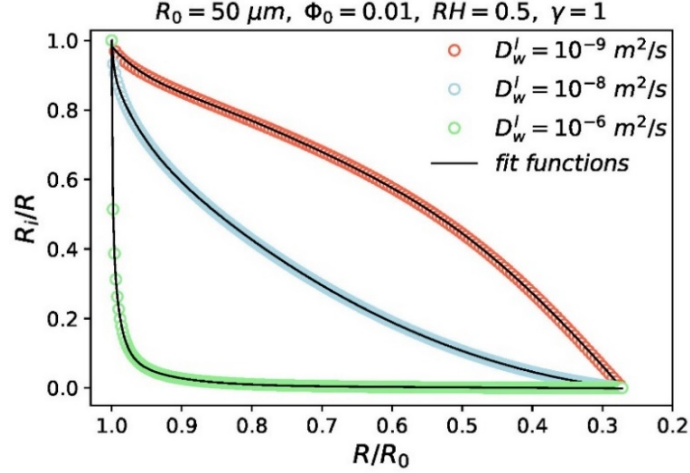

Figure B1 Variation of the radius of the internal core $R_i$ with the droplet radius $R$. The results are calculated from Eq. B30 for $\Phi_0 = 0.01$ and $RH = 0.5$. Three fit functions are introduced and fitted to the data obtained at different internal water diffusion coefficients, depending on which one fits the data best. For example, the results at $D_w^l = 10^{-9}\ m^2/s$ are fitted by $f_1 = \sum_{i=0}^{7} a_i \left(\frac{R}{R_0}\right)^i$, those at $D_w^l = 10^{-8}\ m^2/s$ are fitted by $f_2 = \sum_{i=0}^{3} a_i \left(\frac{R}{R_0}\right)^{b_i}$, and the results at $D_w^l = 10^{-8}\ m^2/s$ are fitted by $f_3 = \sum_{i=0}^{3} a_i \exp\left(\left(\frac{R}{R_0}\right)^{b_i}\right)$.

### B.3.2 The second drying stage

In the second drying stage, $R$ and $c_i$ are considered as time-dependent parameters while $R_i$ is fixed to its value at the end of the first drying stage $R_i^\circ$. Based on equations B24 and B25, therefore, the total evaporation flux and the water concentration profile in the second drying stage can be written as

$$J^{2nd}(t) = \frac{4\pi D_w^l R(t)}{v_w} \left( \frac{1 - \Phi_i(t) - \frac{RH}{\gamma}}{\frac{\alpha}{\gamma} - 1 + \frac{R(t)}{R_i^\circ} + \alpha \varepsilon_T \varepsilon_c (1 - \Phi_i(t))} \right) \tag{B33}$$

$$c_i^{2nd}(r,t) = \begin{cases} \dfrac{1 - \Phi_i(t)}{v_w} & r \leq R_i^\circ \\[1em] \dfrac{1 - \Phi_i(t)}{v_w} + \dfrac{\frac{1}{v_w}\left(1 - \Phi_i(t) - \frac{RH}{\gamma}\right)\left(\frac{R(t)}{r} - \frac{R(t)}{R_i^\circ}\right)}{\frac{\alpha}{\gamma} - 1 + \frac{R(t)}{R_i^\circ} + \alpha \varepsilon_T \varepsilon_c (1 - \Phi_i(t))} & R_i^\circ < r \leq R(t) \end{cases} \tag{B34}$$

$\Phi_i(t)$ in equations B33 and B34 is the time-dependent volume fraction of solutes in the internal core. This parameter can be readily calculated from the solute mass conservation equation (Eq. B29) as

$$\Phi_i = \frac{2\Phi_0 \frac{R_0^3}{R^3}\left(\frac{\alpha}{\gamma} - 1 + \frac{R}{R_i^\circ} + \alpha\varepsilon_T\varepsilon_c\right) - \left(1 - \frac{RH}{\gamma}\right)\left(\frac{R_i^{\circ 2}}{R^2} + 2\frac{R}{R_i^\circ} - 3\right)}{2\left(\frac{\alpha}{\gamma} - 1 + \frac{R}{R_i^\circ} + \alpha\varepsilon_T\varepsilon_c\right) - \left(\frac{R_i^{\circ 2}}{R^2} + 2\frac{R}{R_i^\circ} - 3\right)} \quad (B35)$$

By incorporating $\Phi_i$ from Eq. B35 into equations B33 and B34, one obtains $J^{2nd}$ and $c_i^{2nd}(r)$ at a given time as

$$J^{2nd} = \frac{4\pi D_w^l R}{v_w}\left(\frac{1 - \frac{\Phi_0 R_0^3}{R^3} - \frac{RH}{\gamma}}{\left(\frac{\alpha}{\gamma} - 1 + \frac{R}{R_i^\circ}\right)(1+\mu) + \alpha\varepsilon_T\varepsilon_c\left(1 - \frac{\Phi_0 R_0^3}{R^3} + \mu\frac{RH}{\gamma}\right)}\right) \quad (B36)$$

$$c_i^{2nd}(r) = \begin{cases} \dfrac{1 - \frac{\Phi_0 R_0^3}{R^3} + \mu\frac{RH}{\gamma}}{v_w(1+\mu)} & r \leq R_i^\circ \\[2ex] \dfrac{1 - \frac{\Phi_0 R_0^3}{R^3} + \mu\frac{RH}{\gamma}}{v_w(1+\mu)} + \dfrac{\frac{1}{v_w}\left(1 - \frac{\Phi_0 R_0^3}{R^3} - \frac{RH}{\gamma}\right)\left(\frac{R}{r} - \frac{R}{R_i^\circ}\right)}{\left(\frac{\alpha}{\gamma} - 1 + \frac{R}{R_i^\circ}\right)(1+\mu) + \alpha\varepsilon_T\varepsilon_c\left(1 - \frac{\Phi_0 R_0^3}{R^3} + \mu\frac{RH}{\gamma}\right)} & R_i^\circ < r \leq R \end{cases} \quad (B37)$$

$\mu$ in equations B36 and B37 is a function of $R$, which is given by

$$\mu = -\frac{\frac{R_i^{\circ 2}}{R^2} + 2\frac{R}{R_i^\circ} - 3}{2\left(\frac{\alpha}{\gamma} - 1 + \frac{R}{R_i^\circ} + \alpha\varepsilon_T\varepsilon_c\right)} \quad (B38)$$

Finally, $R(t)$ can be calculated from the total mass conservation of the droplet (Eq. A12), which gives rise to the differential equation

$$\frac{dR}{dt} = -\frac{\frac{D_w^l}{R}\left(1 - \frac{\Phi_0 R_0^3}{R^3} - \frac{RH}{\gamma}\right)}{\left(\frac{\alpha}{\gamma} - 1 + \frac{R}{R_i^\circ}\right)(1+\mu) + \alpha\varepsilon_T\varepsilon_c\left(1 - \frac{\Phi_0 R_0^3}{R^3} + \mu\frac{RH}{\gamma}\right)} \quad (B39)$$

Equation B39 is not analytically solvable and is numerically solved using the Python library "odeint". It is worth noting that here, the droplet radius at the end of the first drying stage $R^\circ$ is used as the initial droplet radius.

Finally, based on the definition of the evaporation time provided in section 2 of the main text (see Eq. 6), the time of the second drying stage is calculated as the time at which the equilibrium droplet radius is 99% of the droplet radius $\tau^{2nd} = t\left(\frac{R_{ev}}{0.99}\right)$.

## B.5 Limit of an infinitely large water diffusion constant in liquid water

An important question here is whether our model recovers the results obtained in the limit of infinitely high internal water diffusivity. From equation B30, we know that in this limit (i.e., when $D_w^l \to \infty$ or, equivalently, $\alpha \to \infty$), $R_i$ goes to zero. This means that the system does not experience the first drying stage and the second stage calculations are valid throughout the evaporation process. In this case, the water concentration profile in the liquid phase can be calculated from Eq. B37 as

$$\lim_{D_w^l \to \infty} c_l(r) = \lim_{\alpha \to \infty} \frac{1 - \frac{\Phi_0 R_0^3}{R^3} + \mu \frac{RH}{\gamma}}{v_w(1+\mu)} + \frac{\frac{1}{v_w}\left(1 - \frac{\Phi_0 R_0^3}{R^3} - \frac{RH}{\gamma}\right)\left(\frac{R}{r} - \frac{R}{R_i^\circ}\right)}{\left(\frac{\alpha}{\gamma} - 1 + \frac{R}{R_i^\circ}\right)(1+\mu) + \alpha\varepsilon_T\varepsilon_c\left(1 - \frac{\Phi_0 R_0^3}{R^3} + \mu\frac{RH}{\gamma}\right)} \qquad (B40)$$

$$= \frac{1}{v_w}\left(1 - \frac{\Phi_0 R_0^3}{R^3}\right)$$

The result equals the water concentration obtained considering the assumption that all particles are evenly distributed in the liquid phase, as expected. One can also obtain the total evaporation flux in the limit of an infinitely large internal water diffusion constant from Eq. B36

$$\lim_{D_w^l \to \infty} J = \lim_{\alpha \to \infty} 4\pi\alpha D_w c_g R \left(\frac{1 - \frac{\Phi_0 R_0^3}{R^3} - \frac{RH}{\gamma}}{\left(\frac{\alpha}{\gamma} - 1 + \frac{R}{R_i^\circ}\right)(1+\mu) + \alpha\varepsilon_T\varepsilon_c\left(1 - \frac{\Phi_0 R_0^3}{R^3} + \mu\frac{RH}{\gamma}\right)}\right) \qquad (B41)$$

$$= 4\pi D_w c_g \gamma R \left(\frac{1 - \frac{\Phi_0 R_0^3}{R^3} - \frac{RH}{\gamma}}{1 + \gamma\varepsilon_T\varepsilon_c\left(1 - \frac{\Phi_0 R_0^3}{R^3}\right)}\right)$$

which results in the evaporation flux obtained without considering the evaporation-induced concentration gradient (see Eq. A11), as expected.

## Appendix C: Dependence of the internal diffusivity profile on the solute concentration profile

In the previous sections, the diffusion equation was solved with the assumption of constant internal water diffusion coefficient. In the presence of solutes, however, this parameter might vary with solute concentration and, consequently, with both time and distance from the droplet center. To find out how such variations affect the evaporation process, we repeat our calculations using a concentration-dependent water diffusion coefficient in the liquid phase. The calculations are performed for the first drying stage, which is the dominant stage when the internal water diffusivity is finite (see Fig. 5a of the main text).

Again, the water concentration profile in the outer shell (i.e., the region that exhibits a concentration gradient) is calculated from the spherical form of the diffusion equation (Eq. B1).

Here, we assume that the internal water diffusion coefficient is linearly dependent on the solute concentration, which approximates the behavior of NaCl and NaBr salts rather accurately [30]. So, we can write

$$D_w^{sol}[c_s(r,t)] = D_w^l(1 - \beta c_s(r,t)) = D_w^l\left(1 - \frac{\beta}{v_s}(1 - v_w c_l(r,t))\right) \quad \text{(C1)}$$

with $D_w^{sol}$ and $D_w^l$ being the water diffusion coefficients in the water solution and in pure water, respectively, and $\beta$ being a constant, which is dependent on the type of solutes. It is known [30] that the water diffusivity decreases in the presence of strongly hydrated solutes, such as NaCl. Therefore, $\beta$ is positive for a solution containing such solutes. The opposite is true for weakly hydrated solutes, such as CsI. For example, the experimental data gives $\beta \simeq 0.065\ M^{-1}$ for NaCl solution [30]. After inserting $D_w^{sol}$ from Eq. C1 into the diffusion equation (Eq. B1), the steady-state form of this equation can be written as

$$\frac{\partial}{\partial r}\left(\left(1 - \frac{\beta}{v_s}(1 - v_w c_l(r))\right)r^2 \frac{\partial c_l(r)}{\partial r}\right) = 0 \quad \text{(C2)}$$

By integrating equation C2 twice and using the two boundary conditions introduced in section B.1 (i.e. $j(R^-) = j(R^+)$ and $c_l(R_i^-) = c_l(R_i^+)$), the water concentration profile inside the droplet can be derived as

$$c_l(r,t) = \begin{cases} \dfrac{1-\Phi_0}{v_w} & r < R_i \\[6pt] \dfrac{1}{v_w} - \dfrac{v_s}{\beta v_w}\left(1 - \sqrt{\left(\dfrac{D_w^{sol,i}}{D_w^l}\right)^2 + \dfrac{2\beta v_w D_w c_0 b_1}{v_s D_w^l}\left(\dfrac{1}{r} - \dfrac{1}{R_i}\right)}\right) & R_i < r < R \end{cases} \quad \text{(C3)}$$

where $D_w^{sol,i} = D_w^l(1 - \frac{\beta}{v_s}(1 - v_w c_i))$ is the initial water diffusion coefficient in the liquid solution. Again, the reactive boundary condition at the droplet surface (Eq. B7) can be used to calculate the unknown parameter $b_1$ in equation C3. This boundary condition gives rise to the equation

$$\frac{RH\beta b_1}{R^2}(D_w + k_c R) + k_c\left(\gamma v_s\left(1 - \frac{\beta}{v_s}\right) + \beta RH\right) = \gamma v_s k_c \sqrt{\left(\frac{D_w^{sol,i}}{D_w^l}\right)^2 + \frac{2\beta v_w D_w c_0 b_1}{v_s D_w^l}\left(\frac{1}{R} - \frac{1}{R_i}\right)} \quad \text{(C4)}$$

In the diffusion-limited regime defined by $D_w < k_c R$, $D_w$ in the left side of Eq. C4 can be safely neglected, resulting in

$$\frac{RH\beta b_1}{R\gamma v_s} + \frac{D_w^{sol,ev}}{D_w^l} = \sqrt{\left(\frac{D_w^{sol,i}}{D_w^l}\right)^2 + \frac{2\beta v_w D_w c_0 b_1}{v_s D_w^l}\left(\frac{1}{R} - \frac{1}{R_i}\right)} \quad \text{(C5)}$$

with $D_w^{sol,ev} = D_w^l(1 - \frac{\beta}{v_s}(1 - RH/\gamma))$ being the internal water diffusion coefficient in the equilibrium state. Equation C5 gives rise to the quadratic equation

$$\zeta^2 + 2\left(\frac{D_w^{sol,ev}}{D_w^l} - \frac{\gamma}{\alpha}\left(1 - \frac{R}{R_i}\right)\right)\zeta + \left(\frac{D_w^{sol,ev}}{D_w^l}\right)^2 - \left(\frac{D_w^{sol,i}}{D_w^l}\right)^2 = 0 \quad (C6)$$

where $\zeta$ is given by

$$\zeta = \frac{RH\beta}{R\gamma v_s} b_1 \quad (C7)$$

From Eq. C6, $\zeta$ can be calculated as

$$\zeta(R, R_i) = -\frac{D_w^{sol,ev}}{D_w^l} + \frac{\gamma}{\alpha}\left(1 - \frac{R}{R_i}\right) + \sqrt{\frac{\gamma^2}{\alpha^2}\left(1 - \frac{R}{R_i}\right)^2 - \frac{2\gamma D_w^{sol,ev}}{\alpha D_w^l}\left(1 - \frac{R}{R_i}\right) + \left(\frac{D_w^{sol,i}}{D_w^l}\right)^2} \quad (C8)$$

By incorporating $b_1$ from Eq. C7 into Eq. C3, the water concentration profile in the liquid phase can be expressed in terms of $\zeta$ as

$$c_l(r) = \begin{cases} \dfrac{1 - \Phi_0}{v_w} & r < R_i \\[2mm] \dfrac{1}{v_w} - \dfrac{v_s}{\beta v_w}\left(1 - \sqrt{\left(\dfrac{D_w^{sol,i}}{D_w^l}\right)^2 + \dfrac{2\gamma}{\alpha}\zeta(R, R_i)\left(\dfrac{R}{r} - \dfrac{R}{R_i}\right)}\right) & R_i < r < R \end{cases} \quad (C9)$$

Also, one can obtain the total water evaporation flux as

$$J = 4\pi R^2 j = 4\pi D_w b_1 c_0 = \frac{4\pi\gamma D_w c_g v_s R}{\beta}\zeta(R, R_i) = \frac{4\pi\gamma D_w^l v_s R}{\alpha\beta v_w}\zeta(R, R_i) \quad (C10)$$

It can be easily shown that in the limit of $\beta = 0$, Eq. C10 results in the water evaporation flux obtained without considering the concentration-dependence of the internal diffusion coefficient (Eq. B21), as expected.

In the next step, the mass conservation equations are employed to determine $R_i$ as a function of $R$, and $R$ as a function of $t$. From the mass conservation of solutes (Eq. B29), one obtains

$$3\int_{R_i}^{R}\left(\sqrt{\left(\frac{D_w^{sol,i}}{D_w^l}\right)^2 + \frac{2\gamma}{\alpha}\zeta(R, R_i)\left(\frac{R}{r} - \frac{R}{R_i}\right)}\right) r^2 dr = \frac{\beta\Phi_0}{v_s}(R_i^3 - R_0^3) + R^3 - R_i^3 \quad (C11)$$

Equation C11 is not analytically solvable and is, therefore, numerically solved. For this, decreasing $R$ and $R_i$ from $R_0$ to 0, the two sides of Eq. C11 are calculated (the left side is calculated through numerical integration), and the pairs at which the above equality is established are kept in two arrays. Then, a polynomial is fitted to the resulting data (similar to what is shown in Fig. B1).

Total mass conservation of the droplet (Eq. A12) gives rise to the differential equation

$$\frac{dR}{dt} = -\frac{D_w^l v_w \gamma}{\beta \alpha R} \left( -\frac{D_w^{sol,ev}}{D_w^l} + \frac{\gamma}{\alpha}\left(1 - \frac{R}{R_i}\right) \right. \tag{C12}$$

$$\left. + \sqrt{\frac{\gamma^2}{\alpha^2}\left(1 - \frac{R}{R_i}\right)^2 - \frac{2\gamma D_w^{sol,ev}}{\alpha D_w^l}\left(1 - \frac{R}{R_i}\right) + \left(\frac{D_w^{sol,i}}{D_w^l}\right)^2} \right)$$

By incorporating the polynomial $R_i(R)$, which is obtained from Eq. C11, into Eq. C12, this equation is numerically solved (again, using Python library "odeint") to calculate $t(R)$. Finally, the evaporation time is obtained as $\tau_{ev} = t(\frac{R_{ev}}{0.99})$, as defined in Eq. 6 of the main text.

## C.2. Evaporation cooling effects

In the last step, our calculations in section C.1 are repeated, this time considering evaporation-induced cooling effects. For this, we go through the same steps as in section A. From the energy balance in the stationary state (Eq. A8) and after replacing $c_g$ in Eq. C10 by $c_g^{surf} = c_g(1 - \varepsilon_c \Delta T)$ (see Eq. A6), one obtains

$$\frac{\beta \Delta T}{\gamma v_s \varepsilon_T (1 - \varepsilon_c \Delta T)} + 1 - \frac{\beta}{v_s} + \frac{\beta RH}{\gamma v_s (1 - \varepsilon_c \Delta T)} - \frac{\gamma(1 - \varepsilon_c \Delta T)}{\alpha}\left(1 - \frac{R}{R_i}\right)$$

$$- \left[\frac{\gamma^2(1 - \varepsilon_c \Delta T)^2}{\alpha^2}\left(1 - \frac{R}{R_i}\right)^2\right. \tag{C13}$$

$$\left. - \frac{2\gamma(1 - \varepsilon_c \Delta T)}{\alpha}\left(1 - \frac{\beta}{v_s} + \frac{\beta RH}{\gamma v_s (1 - \varepsilon_c \Delta T)}\right)\left(1 - \frac{R}{R_i}\right) + \left(\frac{D_w^{sol,i}}{D_w^l}\right)^2\right]^{\frac{1}{2}} = 0$$

Equation C13 is numerically solved to obtain $\Delta T$ as a function of $R_i/R$. For this, decreasing $R_i/R$ from 1 to 0 and $\Delta T$ from its value for a pure water droplet $19.9K(1 - RH)$ [6] to zero, we calculate the two sides of this equation, and keep the pairs at which the above equality is established in two arrays. Then, the polynomial $f = \sum_{i=0}^{7} a_i(R_i/R)^i$ is fitted to the resulting data set to yield $\Delta T(R_i/R)$.

After incorporating $\Delta T(R_i/R)$ in the solute mass conservation equation (Eq. B29), one obtains

$$3\int_{R_i}^{R}\left(\sqrt{\left(\frac{D_w^{sol,i}}{D_w^l}\right)^2 + \frac{2\beta \Delta T(R_i/R)}{\alpha v_s \varepsilon_T}\left(\frac{R}{r} - \frac{R}{R_i}\right)}\right)r^2 dr = \frac{\beta \Phi_0}{v_s}(R_i^3 - R_0^3) + R^3 - R_i^3 \tag{C14}$$

The same numerical method as that explained in section C.1 is used here to calculate $R_i(R)$ from Eq. C14. Then, the functions obtained for $\Delta T(R_i/R)$ and $R_i(R)$ are combined to yield $\Delta T(R)$. Having this function, one can write the total mass conservation equation (Eq. A12) as

$$\frac{dR}{dt} = -\frac{D_w^l}{R\alpha\varepsilon_T}\Delta T(R) \tag{C15}$$

Eq. C15 is numerically solved using Python library "odeint" to calculate $t(R)$ and, consequently, $\tau_{ev} = t(\frac{R_{ev}}{0.99})$.

Finally, the water concentration profile can be expressed according to Eq. C3 as

$$c_l(r) = \begin{cases} \dfrac{1-\Phi_0}{v_w} & r < R_i \\[2ex] \dfrac{1}{v_w} - \dfrac{v_s}{\beta v_w}\left(1 - \sqrt{\left(\dfrac{D_w^{sol,i}}{D_w^l}\right)^2 + \dfrac{2\beta}{\alpha v_s \varepsilon_T}\Delta T(R)\left(\dfrac{R}{r} - \dfrac{R}{R_i}\right)}\right) & R_i < r < R \end{cases} \tag{C16}$$

Equation C16 is used to obtain the solute concentration profile in each step of the numerical solution, which is needed to identify the radius at which the solute concentration at the droplet surface reaches its maximum possible value and a crust forms on the surface (as explained in section 2 of the main text).